\documentclass[12pt,a4paper,titlepage]{article}
\pdfoutput=1
\usepackage[utf8x]{inputenc}
\usepackage
[english]
{babel}

\usepackage{ucs}
\usepackage{amsmath}
\usepackage{amsfonts}
\usepackage{amssymb}
\usepackage{eucal}
\usepackage{mathrsfs}         
\usepackage[sans]{dsfont}     
\usepackage{slashed}
\usepackage[Symbol]{upgreek}  
\usepackage{array}            
\usepackage{longtable}
\usepackage{booktabs}         
\usepackage{colortbl}         
\usepackage{fancyhdr}
\usepackage{multirow}
\usepackage{feynmf}           

\usepackage[
bookmarks]{hyperref}
\usepackage{color}
\hypersetup{colorlinks, bookmarksopen, bookmarksnumbered,
citecolor=verdeCP3, linkcolor=bluCP3, pdfstartview=FitH, urlcolor=rossoCP3
}
\usepackage{color}
\definecolor{grigio}{cmyk}{0,0,0,0.1}
\definecolor{rosa}{cmyk}{0,0.1,0.1,0.02}
\definecolor{rosino}{cmyk}{0,0.05,0.05,0.02}
\definecolor{rosas}{cmyk}{0,0.3,0.25,0.05}
\definecolor{celeste}{cmyk}{0.1,0,0,0.02}
\definecolor{giallino}{cmyk}{0,0,0.1,0.02}
\definecolor{rosso}{cmyk}{0,1,1,0.4}
\definecolor{rossos}{cmyk}{0,1,1,0.55}
\definecolor{rossoc}{cmyk}{0,1,1,0.2}
\definecolor{blu}{cmyk}{1,1,0,0.3}
\definecolor{blus}{cmyk}{1,1,0,0.5}
\definecolor{bluc}{cmyk}{1,1,0,0.1}
\definecolor{blucc}{cmyk}{0.7,0.5,0,0}
\definecolor{viola}{cmyk}{0,1,0,0.6}
\definecolor{viola2}{cmyk}{0,1,0.2,0.6}
\definecolor{verde}{cmyk}{0.92,0,0.59,0.25}
\definecolor{verdec}{cmyk}{0.92,0,0.59,0.15}
\definecolor{verdes}{cmyk}{0.92,0,0.59,0.4}
\definecolor{verdino}{cmyk}{0.12,0,0.09,0.02}
\definecolor{giallo}{cmyk}{0,0,1,0}
\definecolor{gialloverde}{cmyk}{0.44,0,0.74,0}

\definecolor{rossoMIO}{cmyk}{0,.92,.87,.29}
\definecolor{blucMIO}{cmyk}{.7,0,0,0}
\definecolor{violetto}{cmyk}{.31,.31,0,.26}
\definecolor{purple}{cmyk}{.05,.42,0,.30}

\definecolor{rossoCP3}{cmyk}{0,.88,.77,.40}
\definecolor{verdeCP3}{rgb}{0.09765625, 0.57421875, 0.1015625}
\definecolor{bluCP3}{rgb}{0, 0.23, 0.67}

\definecolor{Titolo}{rgb}{0.752941176,0.576470588,0.992156863}
\definecolor{altro}{rgb}{0.094117647,0.650980392,0.643137255}
\definecolor{Peanuts}{rgb}{0.2, 0.4, 0.6}
\definecolor{Pean1}{rgb}{0.6, 0.8, 0.4}
\definecolor{BHO}{rgb}{0.2, 0.8, 1}
\definecolor{Daria}{rgb}{0, 0.9412, 0}
\definecolor{UniPi}{rgb}{0.2549, 0.4627, 0.6275}
\definecolor{UniPidue}{rgb}{0.3216, 0.5804, 0.7882}
\definecolor{verdearXiv}{cmyk}{1,0,1,0}
\definecolor{itemcolor}{cmyk}{.72,.72,0,.30}

\newcommand{\pnt}{\rule[-2mm]{0mm}{7mm}}
\newcommand{\virg}[1]{`#1'}

\newcommand{\wis}[1]{{\mathfrak{v}^#1}^\mu_\nu}
\newcommand{\dwis}[2]{{\mathfrak{w}_{#1}^#2}_\nu}

\newcommand{\doubline}{\hline\addlinespace[0.7mm]\hline}
\newcommand{\nnn}[1]{\boldsymbol{#1}}
\newcommand{\covD}{\mathscr{D}}
\newcommand{\beq}{\begin{equation}}
\newcommand{\eeq}{\end{equation}}

\usepackage
{graphicx}

\long\def\symbolfootnote[#1]#2{\begingroup\def\thefootnote{\fnsymbol{footnote}}
\footnote[#1]{#2}\endgroup}

\begin{document}
\begin{titlepage}
\begin{flushleft}\includegraphics[width=4.5cm]{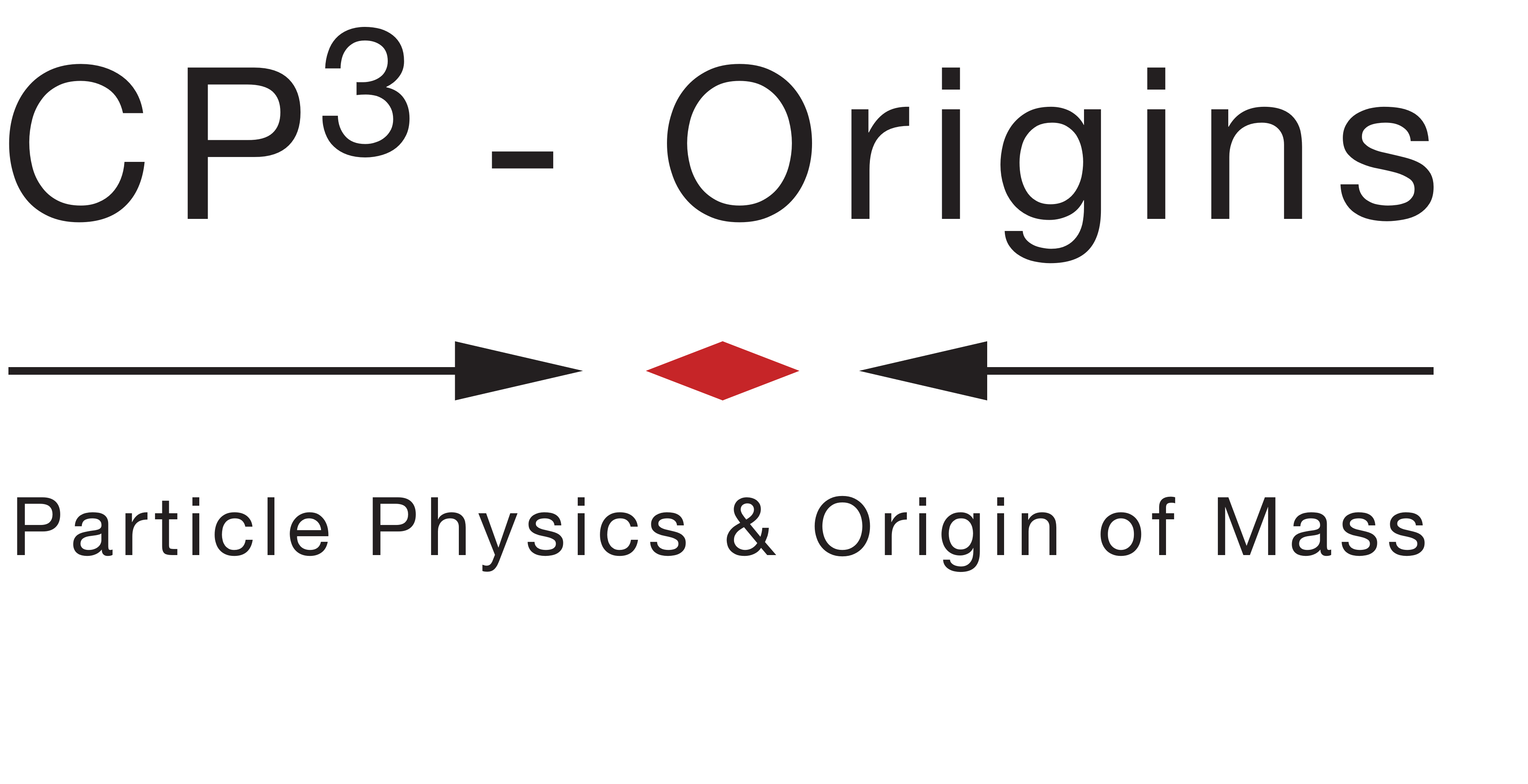} 
\end{flushleft}
 \begin{center}
{{\LARGE {\color{rossoCP3}
\bf  Dark Matter Effective Theory  } 
 }} 
 \end{center}
 \par \vskip .2in \noindent
\begin{center}
{  \Large {\color{black}Eugenio Del Nobile\symbolfootnote[2]{delnobile@cp3-origins.net} \& Francesco Sannino\symbolfootnote[3]{sannino@cp3-origins.net}}  }
\end{center}
\begin{center}
  \par \vskip .1in \noindent
\mbox{\large \it
CP$\,^3$-Origins,
Univ. $\!\!$of Southern Denmark,
Odense, Denmark}
   \par \vskip .1in \noindent
\end{center}
\begin{center}{\large Abstract}\end{center}
\begin{quote}
We organize the effective (self)interaction terms for complex scalar dark matter candidates which are either an isosinglet, isodoublet or an isotriplet with respect to the weak interactions. The classification has been performed ordering the operators in inverse powers of the dark matter cutoff scale. We assume Lorentz invariance, color and charge neutrality. We also introduce potentially interesting dark matter induced flavor-changing operators. Our general framework allows for model independent investigations of dark matter properties.
 \\
 [.1cm]
{
\small \it Preprint: CP$\,^3$-Origins-2011-05}
 \end{quote}
\par \vskip .1in
\vfill
 \end{titlepage}

         \newpage

\def\baselinestretch{1.0}
\tiny
\setlength{\unitlength}{1mm}


\hspace{20mm}
\normalsize

\tableofcontents

\newpage

\section{Introduction}

Arguably models of dynamical electroweak symmetry breaking constitute one of the most natural extensions of the standard model (SM) of particle interactions \cite{Weinberg:1979bn,Susskind:1978ms}. Dark Matter (DM) arises naturally, in these models, as an object made of elementary matter mimicking the bright side of the universe. See \cite{Sannino:2009za} for a review of recent working models of dynamical electroweak symmetry breaking and their DM candidates. DM can arise also in frameworks in which the new dynamics is not directly connected to the breaking of the electroweak scale. All the models can be unified and tested simultaneously when using appropriate low energy effective theory descriptions. 
 
We introduce here a model independent and organized study of the interaction terms with ordinary matter. In this initial classification of the possible terms we consider the following three distinct types of DM: A complex scalar singlet, a doublet and a triplet with respect to weak interactions. The DM candidates are color and electrically neutral scalars charged under a non-SM global $U(1)$ symmetry (e.g. the Technibarion number in Technicolor models). The case of a real scalar subject to a $\mathbb{Z}_2$ symmetry is simply deduced from our list of terms by imposing reality of the scalar field.

The triplet states we considered have the following electric charge assignment, $ T = (T^+, T^0, T^-)$ while the isodoublet is $D = (D^+, D^0)$ or $ (D^0, D^-)$. We indicate the singlet electrically neutral scalar with $\phi$. These states arise, for example, in Technicolor models \cite{Foadi:2008qv, Ryttov:2008xe,Frandsen:2009mi,Gudnason:2006yj,Kainulainen:2006wq,Kouvaris:2007iq,Khlopov:2007ic,Kouvaris:2008hc,Nardi:2008ix}.
\\

This work sets the stage for model independent analyses of DM types and their physical properties. We started with scalars but our approach will be generalized to fermionic type DM in the near future. The various DM effective theories we introduce permit also model independent analyses of DM-genesis were its relic density of symmetric\footnote{By symmetric we mean that it is due to a thermal freeze-out.}, asymmetric or mixed origin \cite{Belyaev:2010kp}.
 
This article is organized as follows. We start with classifying the interaction terms of the complex SM singlet scalar state with ordinary matter. The complete classification, up to (in mass) dimension six operators, of the independent interaction terms with the SM gauge bosons is reported in Appendix \ref{singInt}. In the main text we show only a few terms, some of which already appeared in model dependent phenomenological applications \cite{Foadi:2008qv, Ryttov:2008xe,Frandsen:2009mi}. We also classify here the interactions of the complex neutral scalar with the Higgs in the unitary gauge up to dimension six operators. We show that a further suppression in the ratio of the scalar mass over the scale of the DM physics arises when the latter emerges as a pseudo Goldstone boson of some higher symmetry. This is exactly the case of the Ultra Minimal Technicolor (UMT) model introduced in \cite{Ryttov:2008xe}. Still in the main text we provide a complete classification of the interaction terms with SM fermions including potentially interesting DM induced flavor-changing operators. 

We perform a similar analysis for the doublet and triplet complex scalar DM fields. The summary of the interactions with the SM gauge bosons can be found respectively in Appendices \ref{doubInt} and \ref{tripInt}.

Finally, in Section \ref{Application} we provide an application of the terms introduced in the case of the DM singlet to direct detection searches.

\section{Construction of the Effective Operators}

We used the following standard strategy for constructing the effective operators:

\begin{itemize}
\item Identify the correct degrees of freedom, here the SM and DM fields. 

\item Use the intact global and gauge symmetries to classify the operators. 

\item Provide a counting scheme. 
\end{itemize}

Using the rules above we introduced all possible, to the best of our knowledge, operators ordered in the inverse powers of the cutoff scale $\Lambda$. This scale is assumed to be the one above which a more fundamental theory of DM emerges.  

At low energies the $SU(2)_\text{L} \times U(1)_Y$ symmetry is broken to the electromagnetic one and therefore we classify the operators requiring invariance under color and electromagnetic interactions. Of course, the weak interactions are not violated given that we use the physical eigenstates for the SM fields and the measured couplings among the SM fields. One can also construct the effective Lagrangian directly at the electroweak scale but allowing all allowed interactions (not only the simplest gauge interactions) compatible with the electroweak symmetry. This alternative procedure leads to the same set of operators at low energy given that the relevant intact gauge symmetries are respected.

\section{Singlet}
A scalar SM singlet is one of the most used templates for models of DM. It emerges in a plethora of more or less natural models. It is for this reason that we start our analysis from this DM prototype. We are interested in providing a complete model independent classification of the interaction terms of a generic complex scalar\footnote{Real scalar DM candidates with $\mathbb{Z}_2$ symmetry are easily recovered in our framework by breaking the $U(1)$ global symmetry down to $\mathbb{Z}_2$.} with the SM fields. The natural framework for this analysis are effective theories. We, therefore, first identify a new energy scale $\Lambda$ representing the cutoff of our effective theory above which a more fundamental description arises. Since we are interested in low energy phenomena typically associated to DM detection and cosmic ray production, the low energy effective theory we construct respects explicitly only QCD and electromagnetic symmetries.

\subsection{Interaction with SM gauge bosons}\label{gaugebosons}

 In this section we provide the complete list of operators up to dimension four describing the interaction with the SM gauge bosons. For illustration we also provide a partial list of the operators up to dimension six. The complete list of allowed interaction terms up to and including dimension six operators is provided in Appendix \ref{singInt}. The dimension four operators are:
\begin{gather}
\phi^* \phi \, Z^\mu Z_\mu \ ,
\\
\phi^* \phi \, W^{+ \mu} W^-_\mu \ ,
\\
i \, (\phi^* \overleftrightarrow{\partial_\mu} \phi) \, Z^\mu\equiv J_\mu Z^\mu \ ,\label{J_mu}
\\
\phi^* \phi \, (\partial_\mu Z^\mu) \ .
\end{gather}
Gauge invariance forbids operators with mass dimension less than six involving the ordinary photons and gluons. Therefore the first operators emerge at dimension six and are:
 \begin{gather}
\frac{\phi^* \phi}{\Lambda^2} \, F^{\mu\nu} F_{\mu\nu} \ ,
\qquad 
\frac{\phi^* \phi}{\Lambda^2} \, F_{\mu\nu} \tilde{F}^{\mu\nu} \ ,
\end{gather}
for the interactions with the photon and with $\Lambda$ the energy scale at which these interactions are generated. 
\begin{gather}
\frac{\phi^* \phi}{\Lambda^2} \, G^a_{\mu\nu} G_a^{\mu\nu} \ ,
\qquad 
\frac{\phi^* \phi }{\Lambda^2} \, G^{a}_{\mu\nu} \tilde{G}_a^{\mu\nu} \ ,
\end{gather}
for the interactions with the gluons. The interactions with the gluons are expected to dominate with respect to the ones with the photon. Depending on the model, however, the interaction with the gluons can be further suppressed. 
We can also have: 
\begin{equation}\label{photonint}
J_\mu \frac{\partial_\nu F^{\mu\nu}}{\Lambda^2} \ .
\end{equation}
Direct detection experiments are sensitive to these operators. We do not allow the $U(1)$ symmetry acting on $\phi$ to break spontaneously here. 

For each gauge boson field one must multiply by one power of the associated coupling constant the interaction term in which the field appears. For example the operator $\phi^* \phi Z^{\mu}Z_{\mu}$ should be understood as multiplied by $g^2$ with $g$ the weak coupling constant. Besides the SM coupling constants one has also to multiply each term by an independent dimensionless coefficient whose specific value is fixed once the underlying model of DM is specified.

\subsection{Interaction with SM fermions and Dark Matter induced Flavor Changing Operators}\label{contactoperators}

The possible interaction terms between $\phi$ and Dirac spinors $\psi$ up to dimension six are
\begin{align}
& \phi^* \phi \, \bar{\psi} \psi \ , & \phi^* \phi \, \bar{\psi} \gamma^5 \psi \ ,
\\
& \partial_\mu (\phi^* \phi) \, \bar{\psi} \gamma^\mu \psi \ , & \partial_\mu (\phi^* \phi) \, \bar{\psi} \gamma^\mu \gamma^5 \psi \ ,
\\
& J_\mu \, \bar{\psi} \gamma^\mu \psi \ , & J_\mu \, \bar{\psi} \gamma^\mu \gamma^5 \psi \ ,
\\
& \phi^* \phi \, \bar{\psi} \, i \overleftrightarrow{\slashed{\covD}} \psi \ , & \phi^* \phi \, \bar{\psi} \, i \overleftrightarrow{\slashed{\covD}} \gamma^5 \psi \ ,
\end{align}
where $\bar{\psi}$ and $\psi$ are any two SM fermions such that their combination is colorless and electrically neutral. The covariant derivative \mbox{$\overleftrightarrow{\covD_\mu} = \overleftrightarrow{\partial_\mu} - i e Q A_\mu$} introduces also the minimal coupling to the photon; one has to add the color term $- i T^a G^a_	\mu$ when the covariant derivative is applied to quark fields. For instance it is possible to have
\begin{equation}
\phi^* \phi \, \bar{\mu} e \qquad \text{and} \qquad \phi^* \phi \, \bar{u} \overleftrightarrow{\slashed{\covD}} c \ ,
\end{equation}
where the sum over the color index in the second operator is understood. It is clear that the dark sector can break flavor universality and these operators should be included for a true model independent analysis of the experimental constraints. Each operator should be divided by the appropriate power of the cutoff scale $\Lambda$.

\subsection{$\phi$ as a pseudo-Goldstone boson}

An intriguing possibility is that $\phi$ itself is a Goldstone boson emerging from the spontaneous breaking of non-abelian global symmetries such as $SU(4) \rightarrow Sp(4)$ as it is the case in the UMT model. In this case all its non-derivative interactions vanish unless a mass-term, small with respect to the cutoff scale $\Lambda$, is explicitly introduced. We introduce this mass term and the DM becomes a pseudo-Goldstone boson with the non-derivative interactions suppressed by a factor $m_\phi/\Lambda$ per field $\phi$. For example we would have:
\begin{equation}
\phi^* \phi \, W^\pm_{\mu} W^{\mp \mu} \rightarrow \frac{m^2_{\phi}}{\Lambda^2} \phi^* \phi \, W^\pm_{\mu} W^{\mp \mu} \ .
\end{equation}
According to the new counting of the non-derivative operators the leading dimension four operators are now
\begin{equation}
 J_\mu Z^\mu
 \quad {\rm and} \quad
 \phi^* \phi \, (\partial_\mu Z^\mu) \ .
\end{equation}
%
%
%
%

\subsection{DM-Higgs interaction}
In the unitary gauge the interaction terms involving the Higgs ($h$), were it composite or elementary, up to dimension six are: 
\begin{multline}
\phi^* \phi \sum_{n=1}^{4} a_n \, \frac{h^n}{\Lambda^{n-2}} + (\phi^* \phi)^2 \sum_{n=1}^{2} b_n \, \frac{h^n}{\Lambda^{n}} + (\partial^\mu \phi^*) (\partial_\mu \phi) \sum_{n = 1}^{2} c_n \frac{h^n}{\Lambda^n} +
\\
\partial^\mu (\phi^* \phi) (\partial_\mu h) \sum_{n = 0}^{1} d_n \frac{h^n}{\Lambda^{n+1}} + J^\mu (\partial_\mu h) \sum_{n = 0}^{1} e_n \frac{h^n}{\Lambda^{n+1}} + f \, \phi^* \phi \, \frac{(\partial^\mu h) (\partial_\mu h)}{\Lambda^2}
\end{multline}
where the coefficients are dimensionless and real.

\subsection{DM self-interaction}
DM can self-interact and therefore up to dimension six the Lagrangian terms involving only the DM field are:
\begin{multline}
(\partial^\mu \phi^*) (\partial_\mu \phi) - m_\phi^2 \, \phi^* \phi + \sum_{n = 2}^{3} g_n \frac{(\phi^* \phi)^n}{\Lambda^{2 n - 4}} + \frac{k}{\Lambda^2} \, (\partial^\mu \partial_\mu \phi^*) (\partial^\nu \partial_\nu \phi) +
\\
\frac{1}{\Lambda^2} \, (l_1 \, \partial^\mu (\phi^* \phi) \partial_\mu (\phi^* \phi) + l_2 \, \partial^\mu (\phi^* \phi) J_\mu + l_3 \, J^\mu J_\mu)
%
\end{multline}
where the coefficients are dimensionless and real.

\subsection{Interaction with more than one type of SM fields}
We list here the independent interaction terms with more than one type of SM fields up to dimension six.
\\
The terms involving both DM, the SM gauge bosons and the Higgs are:
\begin{gather}
\phi^* \phi \, Z^\mu Z_\mu h \ ,
\\
\phi^* \phi \, Z^\mu Z_\mu h^2 \ ,
\\
\phi^* \phi \, W^{+\mu} W^-_\mu h \ ,
\\
\phi^* \phi \, W^{+\mu} W^-_\mu h^2 \ ,
\\
\partial_\mu (\phi^* \phi) \, Z^\mu h \ ,
\\
\partial_\mu (\phi^* \phi) \, Z^\mu h^2 \ ,
\\
J_\mu Z^\mu h \ ,
\\
J_\mu Z^\mu h^2 \ ,
\\
\phi^* \phi \, (\partial_\mu Z^\mu) h \ ,
\\
\phi^* \phi \, (\partial_\mu Z^\mu) h^2 \ .
\end{gather}
The terms involving both DM and SM fermions plus either a gauge boson or the Higgs are:
\begin{gather}
\phi^* \phi \, \bar{\psi} \gamma^\mu \psi \, Z_\mu \ ,
\\
\phi^* \phi \, \bar{\psi} \gamma^\mu \gamma^5 \psi \, Z_\mu \ ,
\\
\phi^* \phi \, \bar{\psi} \psi \, h \ ,
\\
\phi^* \phi \, \bar{\psi} \gamma^5 \psi \, h \ ,
\end{gather}
where $\bar{\psi}$ and $\psi$ are any two SM Dirac fermions such that their combination is colorless and electrically neutral. Each operator should be divided by the appropriate power of the cutoff scale $\Lambda$.

\vskip 1cm

The full list of Lagrangian terms introduced here constitutes the first and most comprehensive model independent study of the relevant operators allowing a complex scalar DM candidate to interact with ordinary matter. From the large number of terms it is clear that any recent analysis of DM has explored an incredibly tiny portion of the parameter space in the couplings.

%
%
%
%
%
%
%
%
%
%
%
%
%
%
%
%
%

\section{Doublet}
 Here we generalize the analysis above to the case of the isodoublet. We identify the complex scalar weak doublet to be: 
\begin{equation}
D = \left( \begin{array}{c} D^+
\\
D^0 \end{array}\right) \ ,
\end{equation}
with charge $+1$ with respect to a new global $U(1)$ symmetry. We could equally have defined a new weak doublet $D = (D^0, D^-)$ to which we assign the new abelian charge $-1$. We will first assume that only the first doublet is present. Since at very low energy weak-isospin is broken, $D^0$ and $D^+$ split and the operators involving only $D^0$ are identical to the ones obtained earlier by replacing $\phi$ with $D^0$. Therefore we write here only the new operators. 

\subsection{Interaction of the doublet with SM gauge bosons}
 The dimension 4 operators can be divided in two types:
\begin{itemize}
\item[i)] The terms involving both the neutral and charged components of $D$:
\begin{align}
\lambda \, & {D^0}^* D^+ W^-_\mu Z^\mu + \text{ h.c.} \qquad (\lambda \in \mathbb{C}) \ ,
\\
& (\partial^\mu {D^0}^*) \, D^+ W^-_\mu \ ,
\\
& {D^0}^* (\covD^\mu D^+) \, W^-_\mu \ .
\end{align}
These terms are relevant for the production of $D^0$ from scattering of $D^+$ with nuclei or from decays of $D^+$ if the mass difference between the charged and the neutral component is greater than $M_W$.

\item[ii)] The interaction terms between the charged elements of the multiplet:
\begin{gather}
(\covD^\mu {D^+}^*) (\covD_\mu D^+) \ ,
\\
{D^+}^* D^+ Z^\mu Z_\mu \ ,
\\
{D^+}^* D^+ W^{+\mu} W^-_\mu \ ,
\\
\partial_\mu ({D^+}^* D^+) \, Z^\mu \ ,
\\
i \, ({D^+}^* \overleftrightarrow{\covD_\mu} D^+) \, Z^\mu \ .
\end{gather}

$i \, \partial^\mu ({D^+}^* \overleftrightarrow{\covD_\mu} D^+)$ could also be present, but being a total divergence we will ignore it.
\end{itemize}

The complete list up to and including dimension six operators is reported in Appendix \ref{doubInt}.

\subsection{Interaction of the doublet with SM fermions and Dark Matter induced Flavor Changing Operators}

The possible interaction terms between $D^+$ and Dirac
spinors $\psi$ up to dimension six are
\begin{align} \label{DDDD}
& {D^+}^* D^+ \, \bar{\psi} \psi \ , & {D^+}^* D^+ \, \bar{\psi} \gamma^5 \psi \ ,
\\
& \partial_\mu ({D^+}^* D^+) \, \bar{\psi} \gamma^\mu \psi \ , & \partial_\mu ({D^+}^* D^+) \, \bar{\psi} \gamma^\mu \gamma^5 \psi \ ,
\\
& i \, ({D^+}^* \overleftrightarrow{\covD_{\mu}} D^+) \, \bar{\psi} \gamma^\mu \psi \ , & i \, ({D^+}^* \overleftrightarrow{\covD_{\mu}} D^+) \, \bar{\psi} \gamma^\mu \gamma^5 \psi \ ,
\\
& {D^+}^* D^+ \, \bar{\psi} \, i \overleftrightarrow{\slashed{\covD}} \psi \ , & {D^+}^* D^+ \, \bar{\psi} \, i \overleftrightarrow{\slashed{\covD}} \gamma^5 \psi \ , \label{DDDD-end}
\end{align}
where $\bar{\psi}$ and $\psi$ are any two SM fermions such that their combination is colorless and electrically neutral. Each operator should be divided by the appropriate power of the cutoff scale $\Lambda$. For instance it is possible to have, as done for the $\phi$ case.
\begin{equation}
{D^+}^* D^+ \, \bar{\mu} e \qquad \text{and} \qquad {D^+}^* D^+ \, \bar{u} \overleftrightarrow{\slashed{\covD}} c \ ,
\end{equation}
where the sum over the color index in the second operator is understood. The dark sector can break flavor universality and these operators should be included for a true model independent analysis of the experimental constraints.

The possible interaction terms between $D^0$, $D^+$ and the SM Weyl fermions up to dimension six are

\begin{align}
{D^0}^* D^+ \, &{e_\text{L}}_i \nu_j & \partial_\mu ({D^0}^* D^+) \, &{e_\text{L}}_i \sigma^\mu \bar{\nu}_j
\\
{D^0}^* D^+ \, &{e_\text{L}}_i i \sigma^\mu \covD_\mu \bar{\nu}_j & i \, ({D^0}^* \overleftrightarrow{\covD_{\mu}} D^+) \, &{e_\text{L}}_i \sigma^\mu \bar{\nu}_j \ ,
\\
\notag
\\
{D^0}^* D^+ \, &\bar{\nu}_j \bar{e}^c_{\text{R} i} & \partial_\mu ({D^0}^* D^+) \, &\nu_j \sigma^\mu \bar{e}^c_{\text{R} i}
\\
{D^0}^* D^+ \, &\nu_j i \sigma^\mu \covD_\mu \bar{e}^c_{\text{R} i} & i \, ({D^0}^* \overleftrightarrow{\covD_{\mu}} D^+) \, &\nu_j \sigma^\mu \bar{e}^c_{\text{R} i} \ ,
\\
\notag
\\
{D^0}^* D^+ \, &{u^c_\text{R}}_i {d_\text{L}}_j & \partial_\mu ({D^0}^* D^+) \, &{u^c_\text{R}}_i \sigma^\mu \bar{d}^c_{\text{R} j}
\\
{D^0}^* D^+ \, &{u^c_\text{R}}_i i \sigma^\mu \covD_\mu \bar{d}^c_{\text{R} j} & i \, ({D^0}^* \overleftrightarrow{\covD_{\mu}} D^+) \, &{u^c_\text{R}}_i \sigma^\mu \bar{d}^c_{\text{R} j} \ ,
\\
\notag
\\
{D^0}^* D^+ \, &\bar{d}^c_{\text{R} j} \bar{u}_{\text{L} i} & \partial_\mu ({D^0}^* D^+) \, &{d_\text{L}}_j \sigma^\mu \bar{u}_{\text{L} i}
\\
{D^0}^* D^+ \, &{d_\text{L}}_j i \sigma^\mu \covD_\mu \bar{u}_{\text{L} i} & i \, ({D^0}^* \overleftrightarrow{\covD_{\mu}} D^+) \, &{d_\text{L}}_j \sigma^\mu \bar{u}_{\text{L} i} \ ,
\end{align}
where $i$, $j$ are flavor indices, $\nu = \nu_\text{L}, \nu^c_\text{R}$ and $\sigma^\mu = (I, \sigma^i)$. 
Each operator should be divided by the appropriate power of the cutoff scale $\Lambda$. The couplings of these operators can be complex. 

\subsection{Doublet-Higgs interaction}
In the unitary gauge the interaction terms involving the Higgs field as well as the charged component of the DM doublet, up to dimension six, are:
\begin{multline}
{D^+}^* D^+ \sum_{n=1}^{4} a_n \, \frac{h^n}{\Lambda^{n-2}} + ({D^+}^* D^+) \left( {D^0}^* D^0 \sum_{n=1}^{2} b_n \, \frac{h^n}{\Lambda^{n}} + {D^+}^* D^+ \sum_{n=1}^{2} c_n \, \frac{h^n}{\Lambda^{n}} \right) +
\\
(\covD^\mu {D^+}^*) (\covD_\mu D^+) \sum_{n = 1}^{2} d_n \frac{h^n}{\Lambda^n} + e \, {D^+}^* D^+ \, \frac{(\partial^\mu h) (\partial_\mu h)}{\Lambda^2} +
\\
\partial^\mu ({D^+}^* D^+) (\partial_\mu h) \sum_{n = 0}^{1} f_n \frac{h^n}{\Lambda^{n+1}} + i \, ({D^+}^* \overleftrightarrow{\covD_\mu} D^+) \, (\partial_\mu h) \sum_{n = 0}^{1} g_n \frac{h^n}{\Lambda^{n+1}}
\end{multline}
where the coefficients are dimensionless and real.

\subsection{Doublet self-interaction}\label{DoubSelf}
The doublet has an electrically charged component, therefore in its derivative interactions both the partial as well as the covariant derivative appear. This results in an unavoidable entanglement of the doublet's self-interaction terms and electromagnetic interaction terms. Here we show both the interactions with the purpose to list the entire set of self-interaction operators; see the appendices for the list of all the interaction terms with the photon field.
\\
The Lagrangian terms including the doublet's charged component $D^+$ are, up to dimension six:
\begin{multline}\label{DoubSelfeq}
(\covD^\mu {D^+}^*) (\covD_\mu D^+) - m_+^2 {D^+}^* D^+ +
\\
{D^+}^* D^+ \left(k_1 \, {D^+}^* D^+ + k_2 \, {D^0}^* D^0 \right) +
\\
\frac{1}{\Lambda^2} {D^+}^* D^+ \left(l_1 \, {({D^+}^* D^+)}^2 + l_2 \, {D^+}^* D^+ \, {D^0}^* D^0 + l_3 \, {({D^0}^* D^0)}^2 \right) +
\\
\frac{1}{\Lambda^2} \Big(r_1 \, (\covD^\mu {D^+}^*) (\covD^\mu D^+) \, {D^0}^* D^0 + r_2 \, {D^+}^* D^+ \, (\partial^\mu {D^0}^*) (\partial_\mu D^0) +
\\
r_3 \, \partial^\mu ({D^+}^* D^+) \, \partial_\mu ({D^0}^* D^0) + i \, r_4 \, ({D^+}^* \overleftrightarrow{\covD^\mu} D^+) \, \partial_\mu ({D^0}^* D^0) +
\\
r_5 \, \partial^\mu ({D^+}^* D^+) \, J_\mu + i \, r_6 \, ({D^+}^* \overleftrightarrow{\covD^\mu} D^+) \, J_\mu \Big) + \frac{s}{\Lambda^2} \, (\partial^\mu \partial_\mu {D^+}^*) (\partial^\nu \partial_\nu {D^+})
\end{multline}
where $J_\mu = i \, {D^0}^* \overleftrightarrow{\partial_\mu} D^0$ and the coefficients are dimensionless and real.

\subsection{Interaction of the doublet with more than one type of SM fields}
We list here the independent interaction terms introduced by the presence of an electrically positive DM charged state, beside the neutral one, up to dimension six.
\\
The terms involving both the doublet, the SM gauge bosons and the Higgs are:
\begin{gather}
{D^+}^* D^+ Z^\mu Z_\mu h \ ,
\\
{D^+}^* D^+ Z^\mu Z_\mu h^2 \ ,
\\
{D^+}^* D^+ W^{+\mu} W^-_\mu h \ ,
\\
{D^+}^* D^+ W^{+\mu} W^-_\mu h^2 \ ,
\\
\lambda_1 \, {D^0}^* D^+ W^-_\mu Z^\mu h + \text{ h.c.} \ ,
\\
\lambda_2 \, {D^0}^* D^+ W^-_\mu Z^\mu h^2 + \text{ h.c.} \ ,
\\
\partial_\mu ({D^+}^* D^+) \, Z^\mu h \ ,
\\
\partial_\mu ({D^+}^* D^+) \, Z^\mu h^2 \ ,
\\
i \, ({D^+}^* \overleftrightarrow{\covD_\mu} D^+) Z^\mu h \ ,
\\
i \, ({D^+}^* \overleftrightarrow{\covD_\mu} D^+) Z^\mu h^2 \ ,
\\
{D^+}^* D^+ \, (\partial_\mu Z^\mu) \, h \ ,
\\
{D^+}^* D^+ \, (\partial_\mu Z^\mu) \, h^2 \ ,
\\
\lambda_3 \, (\partial^\mu {D^0}^*) D^+ W^-_\mu h + \text{ h.c.} \ ,
\\
\lambda_4 \, (\partial^\mu {D^0}^*) D^+ W^-_\mu h^2 + \text{ h.c.} \ ,
\\
\lambda_5 \, {D^0}^* (\covD^\mu D^+) W^-_\mu h + \text{ h.c.} \ ,
\\
\lambda_6 \, {D^0}^* (\covD^\mu D^+) W^-_\mu h^2 + \text{ h.c.} \ ,
\\
\lambda_7 \, {D^0}^* D^+ (\covD^\mu W^-_\mu) \, h + \text{ h.c.} \ ,
\\
\lambda_8 \, {D^0}^* D^+ (\covD^\mu W^-_\mu) \, h^2 + \text{ h.c.} \ .
\end{gather}
The terms involving both the doublet, the SM fermions and the gauge bosons are:
\begin{gather}
{D^+}^* D^+ \, \bar{\psi} \gamma^\mu \psi \, Z_\mu \ ,
\\
{D^+}^* D^+ \, \bar{\psi} \gamma^\mu \gamma^5 \psi \, Z_\mu \ ,
\\
\lambda_9 \, {D^0}^* D^+ \, \bar{\psi} \gamma^\mu \psi \, W^-_\mu + \text{ h.c.} \ ,
\\
\lambda_{10} \, {D^0}^* D^+ \, \bar{\psi} \gamma^\mu \gamma^5 \psi \, W^-_\mu + \text{ h.c.} \ ,
\\
\lambda_{11} \, {D^0}^* D^+ \, {e_\text{L}}_i \sigma^\mu \bar{\nu}_j \, Z_\mu + \text{ h.c.} \ ,
\\
\lambda_{12} \, {D^0}^* D^+ \, \nu_j \sigma^\mu \bar{e}^c_{\text{R} i} \, Z_\mu + \text{ h.c.} \ ,
\\
\lambda_{13} \, {D^0}^* D^+ \, {u^c_\text{R}}_i \sigma^\mu \bar{d}^c_{\text{R} j} \, Z_\mu + \text{ h.c.} \ ,
\\
\lambda_{14} \, {D^0}^* D^+ \, {d_\text{L}}_j \sigma^\mu \bar{u}_{\text{L} i} \, Z_\mu + \text{ h.c.} \ ,
\end{gather}
while the terms involving both the doublet, the SM fermions and the Higgs are:
\begin{gather}
{D^+}^* D^+ \, \bar{\psi} \psi \, h \ ,
\\
{D^+}^* D^+ \, \bar{\psi} \gamma^5 \psi \, h \ ,
\\
\lambda_{15} \, {D^0}^* D^+ \, {e_\text{L}}_i \nu_j \, h + \text{ h.c.} \ ,
\\
\lambda_{16} \, {D^0}^* D^+ \, \bar{\nu}_j \bar{e}^c_{\text{R} i} \, h + \text{ h.c.} \ ,
\\
\lambda_{17} \, {D^0}^* D^+ \, {u^c_\text{R}}_i {d_\text{L}}_j \, h + \text{ h.c.} \ ,
\\
\lambda_{18} \, {D^0}^* D^+ \, \bar{d}^c_{\text{R} j} \bar{u}_{\text{L} i} \, h + \text{ h.c.} \ .
\end{gather}
Here all the $\lambda$'s are complex coefficients, and $\bar{\psi}$ and $\psi$ are any two SM Dirac fermions such that their combination is colorless and electrically neutral. Each operator should be divided by the appropriate power of the cutoff scale $\Lambda$.

\section{Triplet}
The complex scalar isotriplet allows for new terms beyond the ones we deduce by replacing $D^0$ with $T^0$, $D^+$ with $T^+$ and, of course, keeping in mind that the terms involving only $T^0$ are the ones for $\phi$. The terms involving $T^-$ and $T^0$ are obtained from the ones of the isodoublet by the formal replacement of $D^+$ with $T^-$ and keeping track of the charges for the SM fields.

\subsection{Interaction of the triplet with SM gauge bosons}
As we have done for the other multiplets, in order to keep the main text easier to read, we list here only the new dimension four operator linking the plus with the minus components of $T$:
\begin{equation}
\lambda \, {T^+}^* T^- W^{+\mu} W^+_\mu + \lambda^* \, {T^-}^* T^+ W^{-\mu} W^-_\mu \qquad (\lambda \in \mathbb{C}) \ .
\end{equation}
The complete list up to and including dimension six operators is reported in Appendix \ref{tripInt}.

\subsection{Interaction of the triplet with SM fermions and Dark Matter induced Flavor Changing Operators}
The triplet introduces the following possible interaction terms (up to dimension six) between the DM and the SM fermions we write here as Weyl spinors:
\begin{align}
{T^+}^* T^- &\, \bar{e}_{\text{L} i} \bar{e}_{\text{L} j} \ ,
\\
{T^+}^* T^- &\, {e^c_\text{R}}_i {e^c_\text{R}}_j \ ,
\\
(\covD_\mu {T^+}^*) T^- &\, {e^c_\text{R}}_i \sigma^\mu \bar{e}_{\text{L} j} \ ,
\\
{T^+}^* (\covD_\mu T^-) &\, {e^c_\text{R}}_i \sigma^\mu \bar{e}_{\text{L} j} \ ,
\end{align}
where $i$, $j$ are flavor indices and $\sigma^\mu = (I, \sigma^i)$. Each operator should be divided by the appropriate power of the cutoff scale $\Lambda$. The couplings for these operators can be complex.

\subsection{Triplet-Higgs interaction}
In the unitary gauge the interaction terms involving the Higgs field as well as both the charged component of the DM triplet, up to dimension six, are:
\begin{equation}
{T^+}^* T^+ {T^-}^* T^- \sum_{n=1}^{2} k_n \, \frac{h^n}{\Lambda^{n}}
\end{equation}
where the coefficients $k_n$ are dimensionless and real.

\subsection{Triplet self-interaction}
As for the doublet, the presence of electrically charged components in the triplet causes the electromagnetic covariant derivative to appear, when needed in the interactions terms. Here we summarize the whole list of self-interaction terms up to dimension six which do not include the covariant derivative as well as the terms including up to two covariant electromagnetic derivatives. In the appendices the reader will find the complete list of the interactions with the photon to the same order in the fields.
\\
The Lagrangian terms including both the triplet's charged components $T^+$ and $T^-$ are, up to dimension six:
\begin{multline}\label{TripSelfeq}
a \, {T^+}^* T^+ {T^-}^* T^- + \tilde{a} \, ({T^+}^* T^0 {T^-}^* T^0 + \text{ h.c.}) +
\\
\frac{1}{\Lambda^2} {T^+}^* T^+ {T^-}^* T^- \left( b_1 \, {T^0}^* T^0 + b_2 \, {T^+}^* T^+ + b_3 \, {T^-}^* T^- \right) +
\\
\frac{1}{\Lambda^2} {T^+}^* T^- \Big( \tilde{b}_1 \, {T^0}^* T^+ {T^0}^* T^+ + \tilde{b}_2 \, {T^+}^* T^+ {T^-}^* T^+ +
\\
\tilde{c}_1 \, {T^-}^* T^0 {T^-}^* T^0 + \tilde{c}_2 \, {T^-}^* T^+ {T^-}^* T^- + \text{ h.c.} \Big) +
\\
\frac{1}{\Lambda^2} {T^+}^* T^0 {T^-}^* T^0 \left( \tilde{d}_1 \, {T^0}^* T^0 + \tilde{d}_2 \, {T^+}^* T^+ + \tilde{d}_3 \, {T^-}^* T^- + \text{ h.c.} \right) +
\\
\frac{1}{\Lambda^2} \Big( c_1 \, \partial^\mu ({T^+}^* T^+) \, \partial_\mu ({T^-}^* T^-) + i \, c_2 \, ({T^+}^* \overleftrightarrow{\covD^\mu} T^+) \, \partial_\mu ({T^-}^* T^-) +
\\
i \, c_3 \, \partial^\mu ({T^+}^* T^+) \, ({T^-}^* \overleftrightarrow{\covD^\mu} T^-) + c_4 \, ({T^+}^* \overleftrightarrow{\covD^\mu} T^+) \, ({T^-}^* \overleftrightarrow{\covD^\mu} T^-) +
\\
c_5 \, (\covD^\mu {T^+}^*) (\covD_\mu T^+) \, {T^-}^* T^- + c_6 \, {T^+}^* T^+ \, (\covD^\mu {T^-}^*) (\covD_\mu T^-) \Big) +
\\
\frac{1}{\Lambda^2} \Big( \tilde{e}_1 \, (\partial^\mu T^0) (\partial_\mu T^0) {T^+}^* {T^-}^* + \tilde{e}_2 \, \partial^\mu (T^0 T^0) (\covD_\mu {T^+}^*) {T^-}^* +
\\
+ \tilde{e}_3 \, \partial^\mu (T^0 T^0) {T^+}^* (\covD_\mu {T^-}^*) + \tilde{e}_4 \, T^0 T^0 (\covD^\mu {T^+}^*) (\covD_\mu {T^-}^*) + \text{ h.c.} \Big)
\end{multline}The coefficients are dimensionless and real, apart from the ones denoted by a tilde that are complex.

\subsection{Interaction of the triplet with more than one type of SM fields}
We list here the independent interaction terms introduced by the simultaneous presence of electrically positive and negative charged DM states, up to dimension six:
\begin{gather}
\lambda_1 \, {T^+}^* T^- W^{+\mu} W^+_\mu h + \text{ h.c.} \ ,
\\
\lambda_2 \, {T^+}^* T^- W^{+\mu} W^+_\mu h^2 + \text{ h.c.} \ ,
\\
\lambda_3 \, {T^+}^* T^- \, {e^c_\text{R}}_i \sigma^\mu \bar{e}_{\text{L} j} Z_\mu + \text{ h.c.} \ ,
\\
\lambda_4 \, {T^+}^* T^- \, \nu_j \sigma^\mu \bar{e}_{\text{L} i} \, W^+_\mu + \text{ h.c.} \ ,
\\
\lambda_5 \, {T^+}^* T^- \, {e^c_\text{R}}_i \sigma^\mu \bar{\nu}_j \, W^+_\mu + \text{ h.c.} \ ,
\\
\lambda_6 \, {T^+}^* T^- \, {d^c_\text{R}}_ j \sigma^\mu \bar{u}^c_{\text{R} i} \, W^+_\mu + \text{ h.c.} \ ,
\\
\lambda_7 \, {T^+}^* T^- \, {u_\text{L}}_ i \sigma^\mu \bar{d}_{\text{L} j} \, W^+_\mu + \text{ h.c.} \ ,
\\
\lambda_8 \, {T^+}^* T^- \, \bar{e}_{\text{L} i} \bar{e}_{\text{L} j} \, h + \text{ h.c.} \ ,
\\
\lambda_9 \, {T^+}^* T^- \, {e^c_\text{R}}_i {e^c_\text{R}}_j \, h + \text{ h.c.} \ .
\end{gather}
All the $\lambda$'s are complex coefficients. Each operator should be divided by the appropriate power of the cutoff scale $\Lambda$.

\section{An application: Direct detection of a DM scalar singlet}\label{Application}

We now provide an example in which we show how to use the operators listed in the previous sections and in the appendices.

In DM direct detection experiments, experimentalists search for nuclear recoils that are compatible with scattering off DM particles. Therefore, in preparing for a phenomenological study of DM direct detection, one has to determine the possible interactions with nucleons. In principle we can have different types of interactions which range from a photon exchange, via for example a tiny dipole (or higher) moment, to an exchange of an Higgs or $Z$ boson, or still another particle, heavier than the DM, for the effective description to be valid. 
\\
The phenomenological relevance of the investigation of these interactions has been put forward in \cite{Foadi:2008qv, DelNobile:2011je}. In these papers it has been shown that the interference of two or more interaction mechanisms can alleviate the tension between the current claims made by different DM direct detection experiments, DAMA/LIBRA  \cite{Savage:2010tg}, CoGeNT \cite{Aalseth:2010vx}, CDMSII  \cite{Ahmed:2010wy}, Xenon 10  \cite{Angle:2011th} and 100 \cite{Aprile:2011hi}. This phenomenon happens if the DM interacts differently with the protons and the neutrons.
\\
The spin-independent cross section for DM scattering off a nucleus with $Z$ protons and $A - Z$ neutrons can be written as
\beq
\sigma = \frac{\mu_A^2}{\pi} \left| Z f_p + (A - Z) f_n \right|^2 = \frac{\mu_A^2}{\mu_p^2} \sigma_p \left| Z + (A - Z) f_n / f_p \right|^2 \ ,
\eeq
where $\mu_A$, $\mu_p$ are respectively the DM-nucleus and DM-proton reduced masses, $\sigma_p = \mu_p^2 \left| f_p \right|^2 / \pi$ is the DM-proton cross section, and $f_n$ and $f_p$ are the overall DM coupling to the neutron and proton \cite{Feng:2011vu}.
\\
A best fit to the data requires $f_n / f_p \simeq -0.71$. Assuming this value for $f_n / f_p$, the fit indicates a DM mass $M_\phi \simeq 8$ GeV and a DM-proton cross section $\sigma_p \simeq 2 \times 10^{-2}$ pb. This corresponds to $| f_p | \simeq 0.14 \times 10^{-5}$ GeV$^{-2}$ and $| f_n | \simeq 0.10 \times 10^{-5}$ GeV$^{-2}$.

Within the framework developed here one can now provide a general interaction amplitude including all of the relevant operators, ordered by power counting. We find useful to distinguish between two different regimes: one in which the dark sector, responsible for the non-renormalizable interactions described in this work, emerges at an energy scale $\Lambda$ lower than the electroweak one, and the other in which this hierarchy is inverted. The difference is that, in the first case, the electroweak gauge bosons and the Higgs field are integrated out together with the unknown dark sector, while in the second case one takes specifically into account the contribution of the $W$ and $Z$ bosons (and of the Higgs, if light enough), and integrates out only the dark sector. Notice that the possible interaction with massless degrees of freedom, such as the photon, is always present regardless of the value of the ultraviolet cut-off.

\subsection{Low energy dark sector}
In the case the dark sector emerges below the electroweak scale, $M_Z > \Lambda \gtrsim M_\phi$, we can describe the DM-nucleon interactions with contact operators. These can be deduced from the ones with quarks listed in Section \ref{contactoperators}; for our purposes the relevant operators are
\beq
\mathscr{L}_\text{contact} = \sum_{N = n, p} \left[ \frac{s_N}{\Lambda} \phi^* \phi \, \bar{N} N + \frac{v_{N1}}{\Lambda^2} \partial_\mu (\phi^* \phi) \, \bar{N} \gamma^\mu N + \frac{v_{N2}}{\Lambda^2} J_\mu \, \bar{N} \gamma^\mu N \right] \ ,
\eeq
where $s$ denotes the scalar couplings to nucleons ($\bar{N} N$) and $v$ the vector ones ($\bar{N} \gamma^\mu N$). We didn't consider pseudo-scalar  ($\bar{N} \gamma^5 N$) and axial-vector couplings ($\bar{N} \gamma^\mu \gamma^5 N$), although they are possible, because their contribution to the cross section is negligible respect to the one given by the operators considered here.
\\
We also consider interaction with the photon; being $\phi$ overall electromagnetically neutral, this interaction can only arise as an effective operator. The most relevant operator for this analysis is the one written in \eqref{photonint}, leading to the Lagrangian term
\beq
\mathscr{L}_\gamma = e \frac{c_\gamma}{\Lambda^2} J_\mu (\partial_\nu F^{\mu\nu}) \ ,
\eeq
where $J_\mu = i \, (\phi^* \overleftrightarrow{\partial_\mu} \phi)$.
\\
Using the effective Lagrangian $\mathscr{L}_\text{contact} + \mathscr{L}_\gamma$, we obtain the following DM couplings to proton and neutron:
\begin{subequations}\label{fpfn}
\begin{align}
f_p &= \frac{s_p}{\Lambda} \frac{1}{2 M_\phi} + \frac{v_{p2}}{\Lambda^2} + 4 \pi \alpha_\text{EM} \frac{c_\gamma}{\Lambda^2} \ ,
\\
f_n &= \frac{s_n}{\Lambda} \frac{1}{2 M_\phi} + \frac{v_{n2}}{\Lambda^2} \ ,
\end{align}
\end{subequations}
$\alpha_\text{EM} = e^2 / 4 \pi$ being the electromagnetic coupling. The $v_{N1}$ couplings don't contribute by virtue of the conservation of the nuclear current $\bar{N} \gamma^\mu N$.
\\
Since $\Lambda \gtrsim M_\phi$, in order to obtain the fitting values for $f_p$ and $f_n$ we need the dimensionless couplings in equations \eqref{fpfn} to be of order $\mathcal{O}(10^{-2}) \div \mathcal{O}(10^{-1})$. Given the large number of parameters involved, it is not an issue to obtain the right value for $f_n / f_p
$.

\subsection{High energy dark sector}

In case the scale of the dark sector, responsible for coupling $\phi$ to the SM, is heavier than or of the order of the electroweak scale, we must consider explicitly the DM couplings to the heavy SM fields. In particular we will consider the case $\Lambda > m_h$, $m_h$ being the physical Higgs mass, that we take here of the order of $100$ GeV. The Lagrangian we must consider is therefore $\mathscr{L}_\text{contact} + \mathscr{L}_\gamma + \mathscr{L}_\text{weak}$, with
\beq
\mathscr{L}_\text{weak} = c_{Z 1} \, J_\mu Z^\mu + c_{Z 2} \, \phi^* \phi \, (\partial_\mu Z^\mu) + c_h v_\text{EW} \, \phi^* \phi \, h \ ,
\eeq
where $v_\text{EW}$ is the Higgs vacuum expectation value. These terms have been classified in Section \ref{gaugebosons}. $\mathscr{L}_\text{contact}$ takes now into account only the physics directly related to the dark sector, without the contribution of the SM fields. Once again we didn't consider pseudo-scalar and axial-vector nucleon couplings, because they give a smaller contribution with respect to the ones considered here. For the same reason we neglect the axial-vector couplings of the $Z$ boson with the quarks in our calculation, and therefore we write the $Z$-nucleon couplings as $i c_{Z, N} \, \bar{N} \gamma^\mu N Z_\mu$, with
\beq
c_{Z, p} = 2 v_{Z, u} + v_{Z, d} \ ,
\qquad
c_{Z, n} = v_{Z, u} + 2 v_{Z, d} \ ,
\eeq
where again $v$ stands for only the vector couplings:
\beq
v_{Z, u} = - \frac{g}{2 \cos\theta_W} \left( \frac{1}{2} - \frac{4}{3} \sin^2 \theta_W \right) \ ,
\qquad
v_{Z, d} = \frac{g}{2 \cos\theta_W} \left( \frac{1}{2} - \frac{2}{3} \sin^2 \theta_W \right) \ .
\eeq
Here we approximated, as it is usual as long as one is concerned only with the vector coupling, the proton $p$ as composed simply by two up quarks $u$ and one down quark $d$, and the neutron $n$ by two $d$'s and one $u$.
\\
We parametrize the Higgs-nucleon coupling as $\frac{m_N}{v_\text{EW}} f \, \bar{N} N \, h$, with $m_N$ the nucleon mass and $f \sim 0.3$ \cite{Shifman:1978zn}.
\\
We find then
\begin{align}
f_p &= - \frac{c_{Z 1} c_{Z, p}}{M_Z^2} + \frac{c_h f m_p}{2 M_\phi m_h^2} + \frac{s_p}{\Lambda} \frac{1}{2 M_\phi} + \frac{v_{p2}}{\Lambda^2} + 4 \pi \alpha_\text{EM} \frac{c_\gamma}{\Lambda^2} \ ,
\\
f_n &= - \frac{c_{Z 1} c_{Z, n}}{M_Z^2} + \frac{c_h f m_n}{2 M_\phi m_h^2} + \frac{s_n}{\Lambda} \frac{1}{2 M_\phi} + \frac{v_{n2}}{\Lambda^2} \ .
\end{align}
As for the couplings $v_{N1}$, also $c_{Z2}$ doesn't contribute to the cross section due to the conservation of the nuclear current.
\\
For a not too heavy dark sector, $1 \text{ TeV} > \Lambda > m_h$, supposing no big difference between the numerical values of the dimensionless couplings,
\begin{align}
f_p &\simeq - \frac{c_{Z 1} c_{Z, p}}{M_Z^2} + \frac{s_p}{\Lambda} \frac{1}{2 M_\phi} \ ,
\\
f_n &\simeq - \frac{c_{Z 1} c_{Z, n}}{M_Z^2} + \frac{s_n}{\Lambda} \frac{1}{2 M_\phi} \ ,
\end{align}
and the couplings need to be of order $\mathcal{O}(10^{-1})$ in order to fit the experimental data. Also in this case it is possible to obtain the fitting value for $f_n / f_p$ (notice that $c_{Z, p}$ and $c_{Z, n}$ have opposite sign, and that also $s_p$, $s_n$ can differ by the sign).
\\
For a heavier dark sector $\Lambda > 1$ TeV, instead,
\begin{equation}
f_p \simeq - \frac{c_{Z 1} c_{Z, p}}{M_Z^2}
\qquad\qquad
f_n \simeq - \frac{c_{Z 1} c_{Z, n}}{M_Z^2} \ ,
\end{equation}
so that if $c_{Z 1} \neq 0$ it is not possible to fit the data, since $f_n / f_p \simeq c_{Z,n} / c_{Z,p} \simeq - 10$. The only possibility is for the DM not to be coupled to the $Z$ boson. In this case, for $\Lambda < 10$ TeV the scattering amplitude due to the contact operator is comparable to the one due to Higgs exchange (still in the assumption that $m_h \sim 100$ GeV), and therefore there is still a possibility to get $f_n / f_p \sim -0.71$. For larger $\Lambda$, instead, getting this value is only possible if also the DM coupling to the Higgs is $0$, i.e.~if the DM couples to quarks only via dark interactions.

\section{Conclusions}

We classified the effective (self)interaction terms for complex scalars assumed to be dark matter candidates which are either an isosinglet, isodoublet or an isotriplet with respect to the weak interactions. The list of the operators has been performed ordering them in the number of derivatives and inverse powers of the DM cutoff scale. We only assumed Lorentz invariance, color and charge neutrality. Our framework allows for future true model independent investigations of dark matter properties.

\section*{Acknowledgments}
We thank Isabella Masina for reading the manuscript and helpful comments.
\newpage

\appendix

\noindent
{\bf \LARGE Appendix}
\noindent
\section{Singlet's Interaction Terms with SM Gauge Bosons}\label{singInt}
 
 We list here all possible hermitian interaction terms of $\phi$ with the SM gauge bosons preserving Lorentz invariance, electric charge and color gauge symmetry up to dimension ($D$) six in the mass. We introduce a number under the column indicated by n which below the table we use to construct explicit relations among the operators. Note that we have not ordered the numbers given that they are just labels. When needed we have also combined operators directly in the table. 

\bigskip

To insure invariance under $U(1)_\text{EM}$ and $SU(3)_\text{c}$ the photon and gluon fields can only enter the Lagrangian in the form of field strength terms and covariant derivatives. Consider that the photon field strength is gauge-neutral and hence its covariant derivative\footnote{We recall that for a generic gauge group the covariant derivative of the field strength $G_{\mu\nu}$ is $\covD_\rho G_{\mu\nu} = \partial_\rho G_{\mu\nu} + \left[ G_\rho, G_{\mu\nu} \right]$ where $G_{\rho}$ is the non-abelian gauge field.
} coincides with its ordinary derivative.
\\

\subsection{Interaction with gluons}
The possible terms involving gluon fields are
\begin{center}
\begin{longtable}{|>{\pnt} r | c | c |}
\hline
n & $D - 4$ & term \\
\hline
& 2 & $\phi^* \phi \, G^a_{\mu\nu} G_a^{\mu\nu}$ \\
\hline
& 2 & $\phi^* \phi \, G^{a}_{\mu\nu} \tilde{G}_a^{\mu\nu}$ \\
\hline
\end{longtable}
\end{center}
where $\tilde{G}_a^{\mu\nu} \equiv \varepsilon^{\mu\nu\rho\sigma} G^a_{\rho\sigma}$ and $G^a_{\rho\sigma}$ is the gluon field strength.

\subsection{Interaction with photons only}
In the same way the possible terms involving only photon fields are
\begin{center}
\begin{longtable}{|>{\pnt} r | c | c |}
\hline
n & $D - 4$ & term \\
\hline
& 2 & $\phi^* \phi \, F^{\mu\nu} F_{\mu\nu}$ \\
\hline
& 2 & $\phi^* \phi \, F_{\mu\nu} \tilde{F}^{\mu\nu}$ \\
\hline
& 2 & $J_\mu (\partial_\nu F^{\mu\nu})$ \\
\hline
\end{longtable}
\end{center}
where
\\
$\tilde{F}^{\mu\nu} \equiv \varepsilon^{\mu\nu\rho\sigma} F_{\rho\sigma}$ , with $F_{\rho\sigma}$ the photon field strength, and
\\
$J_\mu \equiv i \, \phi^* \overleftrightarrow{\partial_\mu} \phi \equiv i \left[ \phi^* (\partial_\mu \phi) - (\partial_\mu \phi^*) \phi \right]$ .

\subsection{Interaction with electroweak gauge bosons $W$, $Z$ and $A$}
We can divide the remaining couplings 
into three classes with regard to the number of derivatives.

\subsubsection{With no derivatives}

\begin{center}
\begin{longtable}{|>{\pnt} r | c | c |}
\hline
n & $D - 4$ & term \\
\hline
& 0 & $\phi^* \phi \, Z^\mu Z_\mu$ \\
\hline
& 0 & $\phi^* \phi \, W^{+ \mu} W^-_\mu$ \\
\hline
& 2 & $\phi^* \phi \, Z^\mu Z_\mu Z^\nu Z_\nu$ \\
\hline
& 2 & $\phi^* \phi \, Z^\mu Z_\mu W^{+ \nu} W^-_\nu$ \\
\hline
& 2 & $\phi^* \phi \, Z^\mu Z^\nu W^+_\mu W^-_\nu$ \\
\hline
& 2 & $\phi^* \phi \, W^{+ \mu} W^-_\mu W^{+ \nu} W^-_\nu$ \\
\hline
& 2 & $\phi^* \phi \, W^{+ \mu} W^+_\mu W^{- \nu} W^-_\nu$ \\
\hline
\end{longtable}
\end{center}

\subsubsection{With one derivative}\label{D^1 phi^2}
We define
\begin{equation}
\wis{\pm} \equiv W^{+\mu} W^{-}_{\nu} \pm W^+_\nu W^{-\mu} \ ,\end{equation}on which the hermitian conjugation acts just exchanging the indices.
\begin{equation}
\dwis{1}{\pm} \equiv (\covD_\mu W^{+\mu}) \, W^{-}_{\nu} \pm W^+_\nu \, (\covD_\mu W^{-\mu}) \ ,
\quad
\dwis{2}{\pm} \equiv W^{+\mu} \, (\covD_\mu W^{-}_{\nu}) \pm (\covD_\mu W^+_\nu) \, W^{-\mu} \ ,
\end{equation}
such that $\wis{+}$, $\dwis{i}{+}$ are hermitian while $\wis{-}$, $\dwis{i}{-}$ are anti-hermitian. These composite operators are electrically neutral, so the superscript sign does not represent their electric charge.
The operators satisfy the following relations:
\begin{equation}
\dwis{1}{\pm} + \dwis{2}{\pm} = \partial_\mu \wis{\pm} \ , \quad \dwis{1}{\pm} - \dwis{2}{\pm} = W^-_\nu \overleftrightarrow{\covD_\mu} W^{+\mu} \mp W^{-\mu} \overleftrightarrow{\covD_\mu} W^+_\nu \ .
\end{equation}
\begin{center}
\begin{longtable}{|>{\pnt} r | c | c |}
\hline
n & $D - 4$ & term \\
\hline
& 0 & $J_\mu Z^\mu$ \\
\hline
& 0 & $\phi^* \phi \, (\partial_\mu Z^\mu)$ \\

\doubline

& 2 & $J_\mu \, Z^\mu Z^{\nu} Z_{\nu}$ \\
\hline
$\nnn{21}$ & 2 & $\partial_\mu (\phi^* \phi) \, Z^\mu Z^{\nu} Z_{\nu}$ \\
\hline
$\nnn{22}$ & 2 & $\phi^* \phi \, (\partial_\mu Z^\mu) \, Z^{\nu} Z_{\nu}$ \\
\hline
$\nnn{23}$ & 2 & $\phi^* \phi \, Z^{\mu} \, (\partial_\mu Z^\nu) \, Z_{\nu}$ \\

\doubline

& 2 & $J_\mu Z^\mu W^{+\nu} W^{-}_{\nu}$ \\
\hline
$\nnn{31}$ & 2 & $\partial_\mu (\phi^* \phi) \, Z^\mu W^{+\nu} W^{-}_{\nu}$ \\
\hline
$\nnn{32}$ & 2 & $\phi^* \phi (\partial_\mu Z^\mu) W^{+\nu} W^{-}_{\nu}$ \\
\hline
$\nnn{33}$ & 2 & $\phi^* \phi \, Z^\mu \partial_\mu (W^{+\nu} W^{-}_{\nu})$ \\
\hline
& 2 & $i \, \phi^* \phi \, Z^\mu \, (W^{-}_{\nu} \overleftrightarrow{\covD_\mu} W^{+\nu})$ \\

\doubline

$\nnn{1}$ & 2 & $i \, [\phi^* (\partial_\mu \phi) \, Z^\nu W^{+\mu} W^{-}_{\nu} - (\partial_\mu \phi^*) \phi \, Z^\nu W^+_\nu W^{-\mu}]$ \\
\hline
$\nnn{2}$ & 2 & $i \, [\phi^* (\partial_\mu \phi) \, Z^\nu W^+_\nu W^{-\mu} - (\partial_\mu \phi^*) \phi \, Z^\nu W^{+\mu} W^-_\nu]$ \\
\hline
$\nnn{1} + \nnn{2}$ & 2 & $J_\mu Z^\nu \wis{+}$ \\
\hline
$\nnn{1} - \nnn{2}$ & 2 & $i \, \partial_\mu (\phi^* \phi) \, Z^\nu \wis{-}$ \\
\hline
$\nnn{3}$ & 2 & $\phi^* (\partial_\mu \phi) \, Z^\nu W^{+\mu} W^{-}_{\nu} + (\partial_\mu \phi^*) \phi \, Z^\nu W^+_\nu W^{-\mu}$ \\
\hline
$\nnn{4}$ & 2 & $\phi^* (\partial_\mu \phi) \, Z^\nu W^+_\nu W^{-\mu} + (\partial_\mu \phi^*) \phi \, Z^\nu W^{+\mu} W^-_\nu$ \\
\hline
$\nnn{3} + \nnn{4}$ & 2 & $\partial_\mu (\phi^* \phi) \, Z^\nu \wis{+}$ \\
\hline
$\nnn{4} - \nnn{3}$ & 2 & $i \, J_\mu Z^\nu \wis{-}$ \\
\hline
$\nnn{9}$ & 2 & $i \, \phi^* \phi \, (\partial_\mu Z^\nu) \wis{-}$ \\
\hline
$\nnn{10}$ & 2 & $\phi^* \phi \, (\partial_\mu Z^\nu) \wis{+}$ \\
\hline
$\nnn{5}$ & 2 & $i \, \phi^* \phi \, Z^\nu \dwis{1}{-}$ \\
\hline
$\nnn{6}$ & 2 & $i \, \phi^* \phi \, Z^\nu \dwis{2}{-}$ \\
\hline
$\nnn{5} + \nnn{6}$ & 2 & $i \, \phi^* \phi \, Z^\nu (\partial_\mu \wis{-})$ \\
\hline
$\nnn{6} - \nnn{5}$ & 2 & $i \, \phi^* \phi \, Z^\nu \, (W^{+\mu} \overleftrightarrow{\covD_\mu} W^-_\nu - W^{-\mu} \overleftrightarrow{\covD_\mu} W^+_\nu)$ \\
\hline
$\nnn{7}$ & 2 & $\phi^* \phi \, Z^\nu \dwis{1}{+}$ \\
\hline
$\nnn{8}$ & 2 & $\phi^* \phi \, Z^\nu \dwis{2}{+}$ \\
\hline
$\nnn{7} + \nnn{8}$ & 2 & $\phi^* \phi \, Z^\nu (\partial_\mu \wis{+})$ \\
\hline
$\nnn{8} - \nnn{7}$ & 2 & $\phi^* \phi \, Z^\nu \, (W^{+\mu} \overleftrightarrow{\covD_\mu} W^-_\nu + W^{-\mu} \overleftrightarrow{\covD_\mu} W^+_\nu)$ \\

\doubline

& 2 & $i \, \varepsilon^{\mu\nu\rho\sigma} J_\mu Z_\nu W^+_{\rho} W^{-}_{\sigma}$ \\
\hline
$\nnn{11}$ & 2 & $i \, \varepsilon^{\mu\nu\rho\sigma} \partial_\mu (\phi^* \phi) \, Z_\nu W^+_{\rho} W^{-}_{\sigma}$ \\
\hline
$\nnn{12}$ & 2 & $i \, \varepsilon^{\mu\nu\rho\sigma} \phi^* \phi \, (\partial_\mu Z_\nu) \, W^+_{\rho} W^{-}_{\sigma}$ \\
\hline
$\nnn{13}$ & 2 & $i \, \varepsilon^{\mu\nu\rho\sigma} \phi^* \phi \, Z_\nu \, \partial_\mu (W^+_{\rho} W^{-}_{\sigma})$ \\
\hline
& 2 & $\varepsilon^{\mu\nu\rho\sigma} \phi^* \phi \, Z_\nu \, (W^+_\rho \overleftrightarrow{\covD_\mu} W^-_\sigma)$ \\

\doubline

& 2 & $i \, \phi^* \phi \, F_{\mu\nu} W^{+\mu} W^{-\nu}$ \\
\hline
& 2 & $i \, \phi^* \phi \, \tilde{F}_{\mu\nu} W^{+\mu} W^{-\nu}$ \\
\hline

\end{longtable}
\end{center}

Not all the operators are independent. In fact the following linear combinations amount to total derivatives:
\\
$\nnn{(3 + 4)} + \nnn{(7 + 8)} + \nnn{10}$
\\
$\nnn{(1 - 2)} + \nnn{(5 + 6)} + \nnn{9}$
\\
$\nnn{11} + \nnn{12} + \nnn{13}$
\\
$\nnn{21} + \nnn{22} + 2 * \nnn{23}$
\\
$\nnn{31} + \nnn{32} + \nnn{33}$.

\subsubsection{With two derivatives}
We can divide these terms in three lists, depending on which gauge boson participates in the interaction term.

\subsubsection{Couplings to a photon and a $Z$ boson}
\begin{center}
\begin{longtable}{|>{\pnt} r | c | c |}
\hline
n & $D - 4$ & term \\
\hline
& 2 & $J_\mu F^{\mu\nu} Z_\nu$ \\
\hline
& 2 & $J_\mu \tilde{F}^{\mu\nu} Z_\nu$ \\
\hline
$\nnn{1}$ & 2 & $\partial_\mu (\phi^* \phi) \, F^{\mu\nu} Z_\nu$ \\
\hline
$\nnn{2}$ & 2 & $\phi^* \phi \, (\partial_\mu F^{\mu\nu}) Z_\nu$ \\
\hline
$\nnn{3}$ & 2 & $\phi^* \phi \, F^{\mu\nu} \, (\partial_\mu Z_\nu)$ \\
\hline
$\nnn{\tilde{1}}$ & 2 & $\partial_\mu (\phi^* \phi) \, \tilde{F}^{\mu\nu} Z_\nu$ \\
\hline
$\nnn{\tilde{3}}$ & 2 & $\phi^* \phi \, \tilde{F}^{\mu\nu} \, (\partial_\mu Z_\nu)$ \\
\hline

\end{longtable}
\end{center}
where the following combinations are total derivatives and therefore vanish:
\\
$\nnn{1} + \nnn{2} + \nnn{3}$
\\
$\nnn{\tilde{1}} + 
\nnn{\tilde{3}}$,
due to the fact that $\partial_\mu \tilde{F}^{\mu\nu} = 0$.

\subsubsection{Couplings to two $Z$ bosons}
\begin{center}
\begin{longtable}{|>{\pnt} r | c | c |}
\hline
n & $D - 4$ & term \\
\hline
$\nnn{1}$ & 2 & $(\partial^\mu \phi^*) (\partial_\mu \phi) \, Z^\nu Z_\nu$ \\
\hline
$\nnn{2}$ & 2 & $[(\partial^\mu \partial_\mu \phi^*) \phi + \phi^* (\partial^\mu \partial_\mu \phi)] \, Z^\nu Z_\nu$ \\
\hline
\multirow{2}{*}{$\nnn{11}$} & \multirow{2}{*}{2} & $- i \, [(\partial^\mu \partial_\mu \phi^*) \phi - \phi^* (\partial^\mu \partial_\mu \phi)] \, Z^\nu Z_\nu$ \\
& & $= (\partial_\mu J^\mu) \, Z^\nu Z_\nu$ \\
\hline
$2 * \nnn{1} + \nnn{2}$ & 2 & $\partial^\mu \partial_\mu (\phi^* \phi) \, Z^\nu Z_\nu$ \\
\hline
$\nnn{3}$ & 2 & $(\partial_\mu \phi^*) (\partial_\nu \phi) \, Z^\mu Z^\nu$ \\
\hline
$\nnn{4}$ & 2 & $[(\partial_\mu \partial_\nu \phi^*) \phi + \phi^* (\partial_\mu \partial_\nu \phi)] \, Z^\mu Z^\nu$ \\
\hline
\multirow{2}{*}{$\nnn{12}$} & \multirow{2}{*}{2} & $- i \, [(\partial_\mu \partial_\nu \phi^*) \phi - \phi^* (\partial_\mu \partial_\nu \phi)] \, Z^\mu Z^\nu$ \\
& & $= (\partial_\mu J_\nu) \, Z^\mu Z^\nu$ \\
\hline
$ 2 * \nnn{3} + \nnn{4}$ & 2 & $\partial_\mu \partial_\nu (\phi^* \phi) \, Z^\mu Z^\nu$ \\
\hline
$\nnn{5}$ & 2 & $\partial^\mu (\phi^* \phi) \, (\partial_\mu Z^\nu) \, Z_\nu$ \\
\hline
$\nnn{6}$ & 2 & $\partial_\mu (\phi^* \phi) \, (\partial_\nu Z^\mu) \, Z^\nu$ \\
\hline
$\nnn{7}$ & 2 & $\partial_\mu (\phi^* \phi) \, (\partial_\nu Z^\nu) \, Z^\mu$ \\
\hline
$\nnn{6} + \nnn{7}$ & 2 & $\partial_\mu (\phi^* \phi) \, \partial_\nu (Z^\mu Z^\nu)$ \\
\hline
$\nnn{7} - \nnn{6}$ & 2 & $\partial_\mu (\phi^* \phi) \, (Z^\mu \overleftrightarrow{\partial_\nu} Z^\nu)$ \\
\hline
$\nnn{8}$ & 2 & $J^\mu \, (\partial_\mu Z^\nu) \, Z_\nu$ \\
\hline
$\nnn{9}$ & 2 & $J_\mu \, (\partial_\nu Z^\mu) \, Z^\nu$ \\
\hline
$\nnn{10}$ & 2 & $J_\mu \, (\partial_\nu Z^\nu) \, Z^\mu$ \\
\hline
$\nnn{9} + \nnn{10}$ & 2 & $J_\mu \, \partial_\nu (Z^\mu Z^\nu)$ \\
\hline
$\nnn{10} - \nnn{9}$ & 2 & $J_\mu \, (Z^\mu \overleftrightarrow{\partial_\nu} Z^\nu)$ \\
\hline
$\nnn{13}$ & 2 & $\phi^* \phi \, (\partial_\mu Z^\mu) \, (\partial_\nu Z^\nu)$ \\
\hline
$\nnn{14}$ & 2 & $\phi^* \phi \, (\partial_\mu Z^\nu) \, (\partial_\nu Z^\mu)$ \\
\hline
$\nnn{15}$ & 2 & $\phi^* \phi \, (\partial^\mu Z^\nu) \, (\partial_\mu Z_\nu)$ \\
\hline
$\nnn{16}$ & 2 & $\phi^* \phi \, (\partial_\mu \partial_\nu Z^\mu) \, Z^\nu$ \\
\hline
$\nnn{17}$ & 2 & $\phi^* \phi \, (\partial^\mu \partial_\mu Z^\nu) \, Z_\nu$ \\
\hline
$2 * (\nnn{15} + \nnn{17})$ & 2 & $\phi^* \phi \, \partial^\mu \partial_\mu (Z^\nu Z_\nu)$ \\
\hline
$\nnn{13} + \nnn{14} +$ & \multirow{2}{*}{2} & \multirow{2}{*}{$\phi^* \phi \, \partial_\mu \partial_\nu (Z^\mu Z^\nu)$} \\
$+ 2 * \nnn{16}$ & & \\
\hline

\end{longtable}
\end{center}
where the following combinations are total derivatives:
\\
$(2 * \nnn{1} + \nnn{2}) + 2 * \nnn{5}$
\\
$(2 * \nnn{3} + \nnn{4}) + (\nnn{6} + \nnn{7})$
\\
$(\nnn{6} + \nnn{7}) + (\nnn{13} + \nnn{14} + 2 * \nnn{16})$
\\
$2 * \nnn{1} + \nnn{2} + 2 * \nnn{5}$
\\
$2 * \nnn{5} + 2 * (\nnn{15} + \nnn{17})$
\\
$\nnn{11} + 2 * \nnn{8}$
\\
$\nnn{12} + (\nnn{9} + \nnn{10})$

\noindent
It is amusing to note that the following relations hold:
\\ $(2 * \nnn{3} + \nnn{4}) + 2 * (\nnn{6} + \nnn{7}) + (\nnn{13} + \nnn{14} + 2 * \nnn{16}) = \partial_\mu \partial_\nu (\phi^* \phi \, Z^\mu Z^\nu)$ \\
and
\\
$2 * \nnn{1} + \nnn{2} + 2 * (2 * \nnn{5}) + 2 * (\nnn{15} + \nnn{17}) = \partial^\mu \partial_\mu (\phi^* \phi \, Z^\nu Z_\nu)$.

\subsubsection{Coupling to two $W$ bosons plus photons}
\begin{center}
\begin{longtable}{|>{\pnt} r | c | c |}
\hline
n & $D - 4$ & term \\
\hline
$\nnn{1}$ & 2 & $(\partial^\mu \phi^*) (\partial_\mu \phi) \, W^{+\nu} W^-_\nu$ \\
\hline
$\nnn{2}$ & 2 & $[(\partial^\mu \partial_\mu \phi^*) \phi + \phi^* (\partial^\mu \partial_\mu \phi)] \, W^{+\nu} W^-_\nu$ \\
\hline
\multirow{2}{*}{$\nnn{11}$} & \multirow{2}{*}{2} & $- i \, [(\partial^\mu \partial_\mu \phi^*) \phi - \phi^* (\partial^\mu \partial_\mu \phi)] \, W^{+\nu} W^-_\nu$ \\
& & $= (\partial_\mu J^\mu) \, W^{+\nu} W^-_\nu$ \\
\hline
$2 * \nnn{1} + \nnn{2}$ & 2 & $\partial^\mu \partial_\mu (\phi^* \phi) \, W^{+\nu} W^-_\nu$ \\
\hline
$\nnn{3a}$ & 2 & $(\partial_\mu \phi^*) (\partial_\nu \phi) \, W^{+\mu} W^{-\nu}$ \\
\hline
$\nnn{3b}$ & 2 & $(\partial_\mu \phi^*) (\partial_\nu \phi) \, W^{+\nu} W^{-\mu}$ \\
\hline
$\nnn{3a} + \nnn{3b}$ & 2 & $(\partial_\mu \phi^*) (\partial^\nu \phi) \, \wis{+}$ \\
\hline
$\nnn{3a} - \nnn{3b}$ & 2 & $(\partial_\mu \phi^*) (\partial^\nu \phi) \, \wis{-} = - \frac{i}{2} \, (\partial_\mu J^\nu) \, \wis{-}$ \\
\hline
$\nnn{4}$ & 2 & $[(\partial_\mu \partial_\nu \phi^*) \phi + \phi^* (\partial_\mu \partial_\nu \phi)] \, W^{+\mu} W^{-\nu}$ \\
\hline
\multirow{2}{*}{$\nnn{12}$} & \multirow{2}{*}{2} & $(\partial_\mu J^\nu) \, \wis{+} =$ \\
& & $- 2 \, i \, [(\partial_\mu \partial_\nu \phi^*) \phi - \phi^* (\partial_\mu \partial_\nu \phi)] \, W^{+\mu} W^{-\nu}$ \\
\hline
$\nnn{3a} + \nnn{3b} + \nnn{4}$ & 2 & $\partial_\mu \partial_\nu (\phi^* \phi) \, W^{+\mu} W^{-\nu}$ \\
\hline
$\nnn{5a}$ & 2 & $\partial^\mu (\phi^* \phi) \, \partial_\mu (W^{+\nu} W^-_\nu)$ \\
\hline
$\nnn{5b}$ & 2 & $i \, \partial^\mu (\phi^* \phi) \, (W^{-\nu} \overleftrightarrow{\covD_\mu} W^+_\nu)$ \\
\hline
$\nnn{6a}$ & 2 & $\partial^\nu (\phi^* \phi) \, \dwis{1}{+}$ \\
\hline
$\nnn{6b}$ & 2 & $i \, \partial^\nu (\phi^* \phi) \, \dwis{1}{-}$ \\
\hline
$\nnn{7a}$ & 2 & $\partial^\nu (\phi^* \phi) \, \dwis{2}{+}$ \\
\hline
$\nnn{7b}$ & 2 & $i \, \partial^\nu (\phi^* \phi) \, \dwis{2}{-}$ \\
\hline
$\nnn{6a} + \nnn{7a}$ & 2 & $\partial^\nu (\phi^* \phi) \, (\partial_\mu \wis{+})$ \\
\hline
$\nnn{6a} - \nnn{7a}$ & 2 & $\partial^\nu (\phi^* \phi) \, (W^-_\nu \overleftrightarrow{\covD_\mu} W^{+\mu} - W^{-\mu} \overleftrightarrow{\covD_\mu} W^+_\nu)$ \\
\hline
$\nnn{6b} + \nnn{7b}$ & 2 & $i \, \partial^\nu (\phi^* \phi) \, (\partial_\mu \wis{-})$ \\
\hline
$\nnn{6b} - \nnn{7b}$ & 2 & $i \, \partial^\nu (\phi^* \phi) \, (W^-_\nu \overleftrightarrow{\covD_\mu} W^{+\mu} + W^{-\mu} \overleftrightarrow{\covD_\mu} W^+_\nu)$ \\

\doubline

$\nnn{8a}$ & 2 & $J^\mu \partial_\mu (W^{+\nu} W^-_\nu)$ \\
\hline
$\nnn{8b}$ & 2 & $i \, J^\mu \, (W^{-\nu} \overleftrightarrow{\covD_\mu} W^+_\nu)$ \\
\hline
$\nnn{9a}$ & 2 & $J^\nu \dwis{1}{+}$ \\
\hline
$\nnn{9b}$ & 2 & $i \, J^\nu \dwis{1}{-}$ \\
\hline
$\nnn{10a}$ & 2 & $J^\nu \dwis{2}{+}$ \\
\hline
$\nnn{10b}$ & 2 & $i \, J^\nu \dwis{2}{-}$ \\
\hline
$\nnn{9a} + \nnn{10a}$ & 2 & $J^\nu \, (\partial_\mu \wis{+})$ \\
\hline
$\nnn{9a} - \nnn{10a}$ & 2 & $J^\nu \, (W^-_\nu \overleftrightarrow{\covD_\mu} W^{+\mu} - W^{-\mu} \overleftrightarrow{\covD_\mu} W^+_\nu)$ \\
\hline
$\nnn{9b} + \nnn{10b}$ & 2 & $i \, J^\nu \, (\partial_\mu \wis{-})$ \\
\hline
$\nnn{9b} - \nnn{10b}$ & 2 & $i \, J^\nu \, (W^-_\nu \overleftrightarrow{\covD_\mu} W^{+\mu} + W^{-\mu} \overleftrightarrow{\covD_\mu} W^+_\nu)$ \\

\doubline

$\nnn{13}$ & 2 & $\phi^* \phi \, (\covD_\mu W^{+\mu}) \, (\covD_\nu W^{-\nu})$ \\
\hline
$\nnn{14}$ & 2 & $\phi^* \phi \, (\covD_\mu W^{+\nu}) \, (\covD_\nu W^{-\mu})$ \\
\hline
$\nnn{15}$ & 2 & $\phi^* \phi \, (\covD^\mu W^{+\nu}) \, (\covD_\mu W^-_\nu)$ \\
\hline
$\nnn{16a}$ & 2 & $\phi^* \phi \, [(\covD_\mu \covD_\nu W^{+\mu}) \, W^{-\nu} + W^{+\mu} \, (\covD_\mu \covD_\nu W^{-\nu})]$ \\
\hline
\multirow{3}{*}{$\nnn{16b}$} & \multirow{3}{*}{2} & $2 \, i \, \phi^* \phi \, [(\covD_\mu \covD_\nu W^{+\mu}) \, W^{-\nu} - W^{+\mu} \, (\covD_\mu \covD_\nu W^{-\nu})]$ \\
& & $= i \, \phi^* \phi \, \partial_\mu (W^{-\nu} \overleftrightarrow{\covD_\nu} W^{+\mu} + W^{-\mu} \overleftrightarrow{\covD_\nu} W^{+\nu})$ \\
& & $= i \, \phi^* \phi \, \partial^\nu (\dwis{2}{-} - \dwis{1}{-})$ \\
\hline
$\nnn{17a}$ & 2 & $\phi^* \phi \, [(\covD^\mu \covD_\mu W^{+\nu}) \, W^-_\nu + W^{+\nu} \, (\covD^\mu \covD_\mu W^-_\nu)]$ \\
\hline
\multirow{2}{*}{$\nnn{17b}$} & \multirow{2}{*}{2} & $i \, \phi^* \phi \, [(\covD^\mu \covD_\mu W^{+\nu}) \, W^-_\nu - W^{+\nu} \, (\covD^\mu \covD_\mu W^-_\nu)]$ \\
& & $= i \, \phi^* \phi \, \partial^\mu (W^{-\nu} \overleftrightarrow{\covD_\mu} W^+_\nu)$ \\
\hline
$2 * \nnn{15} + \nnn{17a}$ & 2 & $\phi^* \phi \, \partial^\mu \partial_\mu (W^{+\nu} W^-_\nu)$ \\
\hline
$\nnn{13} + \nnn{14} + \nnn{16a}$ & 2 & $\phi^* \phi \, \partial_\mu \partial_\nu (W^{+\mu} W^{-\nu})$ \\
\hline
\multirow{2}{*}{$2 * (\nnn{14} - \nnn{13})$} & \multirow{2}{*}{2} & $\phi^* \phi \, \partial_\mu (W^{-\nu} \overleftrightarrow{\covD_\nu} W^{+\mu} - W^{-\mu} \overleftrightarrow{\covD_\nu} W^{+\nu})$ \\
& & $= \phi^* \phi \, \partial^\nu (\dwis{2}{+} - \dwis{1}{+})$ \\

\doubline

$\nnn{21}$ & 2 & $\varepsilon^{\mu\nu\rho\sigma} (\partial_\mu \phi^*) (\partial_\nu \phi) \, W^+_{\rho} W^{-}_{\sigma}$ \\
\hline
& 2 & $i \, \varepsilon^{\mu\nu\rho\sigma} \partial_\mu (\phi^* \phi) \, \partial_\nu (W^+_{\rho} W^{-}_{\sigma}) \xrightarrow{\int \text{by parts}} 0$ \\
\hline
$\nnn{22}$ & 2 & $\varepsilon^{\mu\nu\rho\sigma} \partial_\mu (\phi^* \phi) \, (W^+_{\rho} \overleftrightarrow{\covD_\nu} W^{-}_{\sigma})$ \\
\hline
$\nnn{23}$ & 2 & $i \, \varepsilon^{\mu\nu\rho\sigma} J_\mu \, \partial_\nu (W^+_{\rho} W^{-}_{\sigma})$ \\
\hline
& 2 & $\varepsilon^{\mu\nu\rho\sigma} J_\mu \, (W^+_{\rho} \overleftrightarrow{\covD_\nu} W^{-}_{\sigma})$ \\
\hline
$\nnn{24}$ & 2 & $\varepsilon^{\mu\nu\rho\sigma} \phi^* \phi \, (\partial_\mu W^+_{\rho}) (\partial_\nu W^{-}_{\sigma})$ \\

\hline
\end{longtable}
\end{center}
where the following combinations are total derivatives:
\\
$(2 * \nnn{1} + \nnn{2}) + \nnn{5a}$ and $\nnn{5a} + (2 * \nnn{15} + \nnn{17a})$
\\
$\nnn{11} + \nnn{8a}$
\\
$2 * (\nnn{3a} + \nnn{3b} + \nnn{4}) + (\nnn{6a} + \nnn{7a})$
\\
$\nnn{12} + (\nnn{9a} + \nnn{10a})$
\\
$- 2 * (\nnn{3a} - \nnn{3b}) + (\nnn{9b} + \nnn{10b})$
\\
$\nnn{5b} + \nnn{17b}$
\\
$\nnn{7a} - \nnn{6a} + 2 * (\nnn{14} - \nnn{13})$
\\
$\nnn{7b} - \nnn{6b} + \nnn{16b}$
\\
$(\nnn{6a} + \nnn{7a}) + 2 * (\nnn{13} + \nnn{14} + \nnn{16a})$
\\
We also have 
\\ $(2 * \nnn{1} + \nnn{2}) + 2 * \nnn{5a} + (2 * \nnn{15} + \nnn{17a}) = \partial^\mu \partial_\mu (\phi^* \phi \, W^{+\nu} W^-_\nu)$

\noindent
The epsilon terms enjoy the relations:
\\
$2 * \nnn{21} + \nnn{23}$
\\
$2 * \nnn{24} + \nnn{22}$.

\subsubsection{With three derivatives}

The only possible SM gauge boson that can enter the interaction terms with three derivatives is the $Z$ boson.
\begin{center}
\begin{longtable}{|>{\pnt} r | c | c |}
\hline
n & $D - 4$ & term \\
\hline
$\nnn{1}$ & 2 & $ \phi^* \phi \, (\partial^2 \partial_\mu Z^\mu)$ \\

\doubline

$\nnn{2}$ & 2 & $\partial_\mu (\phi^* \phi) \, (\partial^2 Z^\mu)$ \\
\hline
$\nnn{3}$ & 2 & $\partial^\mu (\phi^* \phi) \, (\partial_\mu \partial_\nu Z^\nu)$ \\
\hline
$\nnn{4}$ & 2 & $J_\mu \, (\partial^2 Z^\mu)$ \\
\hline
$\nnn{5}$ & 2 & $J^\mu \, (\partial_\mu \partial_\nu Z^\nu)$ \\

\doubline

$\nnn{6}$ & 2 & $(\partial^\mu \phi^*) (\partial_\mu \phi) \, (\partial_\nu Z^\nu)$ \\
\hline
$\nnn{7}$ & 2 & $[(\partial_\mu \phi^*) (\partial_\nu \phi) + (\partial_\nu \phi^*) (\partial_\mu \phi)] \, (\partial^\mu Z^\nu)$ \\
\hline
$\nnn{8}$ & 2 & $i \, [(\partial_\mu \phi^*) (\partial_\nu \phi) - (\partial_\nu \phi^*) (\partial_\mu \phi)] \, (\partial^\mu Z^\nu)$ \\
\hline
$\nnn{9}$ & 2 & $i \, \varepsilon^{\mu\nu\rho\sigma} \, (\partial_\mu \phi^*) (\partial_\nu \phi) \, (\partial_\rho Z_\sigma) \xrightarrow{\int \text{by parts}} 0$ \\

\doubline

$\nnn{10}$ & 2 & $[(\partial^2 \phi^*) \phi + \phi^* (\partial^2 \phi)] \, (\partial_\mu Z^\mu)$ \\
\hline
\multirow{2}{*}{$\nnn{11}$} & \multirow{2}{*}{2} & $- i \, [(\partial^2 \phi^*) \phi - \phi^* (\partial^2 \phi)] \, (\partial_\mu Z^\mu)$ \\
& & $= (\partial_\mu J^\mu) \, (\partial_\nu Z^\nu)$ \\
\hline
$\nnn{12}$ & 2 & $[(\partial_\mu \partial_\nu \phi^*) \phi + \phi^* (\partial_\mu \partial_\nu \phi)] \, (\partial^\mu Z^\nu)$ \\
\hline
$\nnn{13}$ & 2 & $i \, [(\partial_\mu \partial_\nu \phi^*) \phi - \phi^* (\partial_\mu \partial_\nu \phi)] \, (\partial^\mu Z^\nu)$ \\
\hline
$2 * \nnn{6} + \nnn{10}$ & 2 & $\partial^\mu \partial_\mu (\phi^* \phi) \, (\partial_\nu Z^\nu)$ \\
\hline
$\nnn{7} + \nnn{12}$ & 2 & $\partial_\mu \partial_\nu (\phi^* \phi) \, (\partial^\mu Z^\nu)$ \\
\hline
$\nnn{8} - \nnn{13}$ & 2 & $(\partial^\mu J_\nu) \, (\partial_\mu Z^\nu)$ \\
\hline
$- (\nnn{8} + \nnn{13})$ & 2 & $(\partial_\mu J^\nu) \, (\partial_\nu Z^\mu)$ \\

\doubline

$\nnn{14}$ & 2 & $[(\partial^2 \phi^*) (\partial_\mu \phi) + (\partial_\mu \phi^*) (\partial^2 \phi)] \, Z^\mu$ \\
\hline
$\nnn{15}$ & 2 & $i \, [(\partial^2 \phi^*) (\partial_\mu \phi) - (\partial_\mu \phi^*) (\partial^2 \phi)] \, Z^\mu$ \\
\hline
\multirow{2}{*}{$\nnn{16}$} & \multirow{2}{*}{2} & $[(\partial_\mu \partial_\nu \phi^*) (\partial^\mu \phi) + (\partial^\mu \phi^*) (\partial_\mu \partial_\nu \phi)] \, Z^\nu$ \\
& & $= \partial_\mu [(\partial^\nu \phi^*) (\partial_\nu \phi)] \, Z^\mu$ \\
\hline
$\nnn{17}$ & 2 & $i \, [(\partial_\mu \partial_\nu \phi^*) (\partial^\mu \phi) - (\partial^\mu \phi^*) (\partial_\mu \partial_\nu \phi)] \, Z^\nu$ \\
\hline
$\nnn{18}$ & 2 & $[(\partial^2 \partial_\mu \phi^*) \phi + \phi^* (\partial^2 \partial_\mu \phi)] \, Z^\mu$ \\
\hline
$\nnn{19}$ & 2 & $i \, [(\partial^2 \partial_\mu \phi^*) \phi - \phi^* (\partial^2 \partial_\mu \phi)] \, Z^\mu$ \\
\hline
$\nnn{14} + \nnn{16}$ & 2 & $\partial^\mu [(\partial_\mu \phi^*) (\partial_\nu \phi) + (\partial_\nu \phi^*) (\partial_\mu \phi)] \, Z^\nu$ \\
\hline
$\nnn{15} - \nnn{17}$ & 2 & $i \, \partial^\mu [(\partial_\mu \phi^*) (\partial_\nu \phi) - (\partial_\nu \phi^*) (\partial_\mu \phi)] \, Z^\nu$ \\
\hline
$\nnn{16} + \nnn{18}$ & 2 & $\partial^\mu [(\partial_\mu \partial_\nu \phi^*) \phi + \phi^* (\partial_\mu \partial_\nu \phi)] \, Z^\nu$ \\
\hline
$\nnn{17} + \nnn{19}$ & 2 & $i \, \partial^\mu [(\partial_\mu \partial_\nu \phi^*) \phi - \phi^* (\partial_\mu \partial_\nu \phi)] \, Z^\nu$ \\
\hline
$(\nnn{14} + \nnn{16}) +$ & \multirow{2}{*}{2} & \multirow{2}{*}{$\partial^2 \partial_\mu (\phi^* \phi) \, Z^\mu$} \\
$(\nnn{16} + \nnn{18})$ & & \\
\hline
$(\nnn{15} - \nnn{17}) -$ & \multirow{2}{*}{2} & \multirow{2}{*}{$(\partial^2 J_\mu) \, Z^\mu$} \\
$(\nnn{17} + \nnn{19})$ & & \\
\hline
$- (\nnn{15} - \nnn{17})$ & \multirow{3}{*}{2} & \multirow{3}{*}{$(\partial_\mu \partial_\nu J^\mu) \, Z^\nu$} \\
$- (\nnn{17} + \nnn{19})$ & & \\
$= - \nnn{15} - \nnn{19}$ & & \\
\hline

\end{longtable}
\end{center}

Not all these terms are independent, due to the vanishing of the following total derivatives:
\\
$\nnn{1} + \nnn{2}$ and $\nnn{1} + \nnn{3}$
\\
$\nnn{6} + \nnn{16}$
\\
$\nnn{7} + (\nnn{14} + \nnn{16})$
\\
$\nnn{12} + (\nnn{16} + \nnn{18})$
\\
$\nnn{13} + (\nnn{17} + \nnn{19})$
\\
$\nnn{2} + (\nnn{7} + \nnn{12})$ and $\nnn{3} + (\nnn{7} + \nnn{12})$ and $\nnn{3} + (2 * \nnn{6} + \nnn{10})$
\\
$(\nnn{7} + \nnn{12}) + (\nnn{14} + \nnn{16}) + (\nnn{16} + \nnn{18})$ and $(2 * \nnn{6} + \nnn{10}) + (\nnn{14} + \nnn{16}) + (\nnn{16} + \nnn{18})$
\\
$\nnn{4} + (\nnn{8} - \nnn{13})$ and $(\nnn{8} - \nnn{13}) + ((\nnn{15} - \nnn{17}) - (\nnn{17} + \nnn{19}))$
\\
$\nnn{5} - (\nnn{8} + \nnn{13})$ and $- (\nnn{8} + \nnn{13}) + (- \nnn{15} - \nnn{19})$
\\
$\nnn{5} + \nnn{11}$ and $\nnn{11} + (- \nnn{15} - \nnn{19})$.

\subsection{Terms with four $\phi$'s}
The possible terms that are quartic in the fields $\phi$ only concern couplings to the weak gauge bosons:
\begin{center}
\begin{longtable}{|>{\pnt} r | c | c |}
\hline
n & $D - 4$ & term \\
\hline
& 2 & ${(\phi^* \phi)}^2 \, Z^\mu Z_\mu$ \\
\hline
& 2 & ${(\phi^* \phi)}^2 \, (\partial_\mu Z^\mu)$ \\
\hline
& 2 & ${(\phi^* \phi)}^2 \, W^{+ \mu} W^-_\mu$ \\
\hline
& 2 & $\phi^* \phi \, J^\mu Z_\mu$ \\
\hline
\end{longtable}
\end{center}

\section{Doublet's Interaction Terms with SM Gauge Bosons}\label{doubInt}

We enlarge the lists provided in Appendix \ref{singInt} to include the interactions of an electrically charged particle $D^\pm$. The neutral state, that we call here $D^0$, has to be identified with $\phi$ in the previous appendix. The two states can be regarded as a weak doublet with hypercharge $Y = \pm 1/2$, also charged under a non-SM global $U(1)$ symmetry (e.g. the Technibarion number). We impose invariance under this extra abelian symmetry at the Lagrangian level.
\\
To obtain the full list of interaction-terms with the SM gauge bosons one should consider the terms presented here and in the previous section. 

We assume the Lagrangian terms to be hermitians, and to preserve Lorentz invariance, electric charge and color gauge symmetry. We include the operators up to dimension six in the mass. For further details and the description of the notation see Appendix \ref{singInt}.

\bigskip

The photon and gluon fields enter the Lagrangian only via their field strengths and covariant derivatives.

%
%

\subsection{Interaction with gluons only}
\begin{center}
\begin{longtable}{|>{\pnt} r | c | c |}
\hline
n & $D - 4$ & term \\
\hline
& 2 & ${D^\pm}^* D^\pm G^a_{\mu\nu} G_a^{\mu\nu}$ \\
\hline
& 2 & ${D^\pm}^* D^\pm G^a_{\mu\nu} \tilde{G}_a^{\mu\nu}$ \\
\hline
\end{longtable}
\end{center}
where $\tilde{G}_a^{\mu\nu} \equiv \varepsilon^{\mu\nu\rho\sigma} G^a_{\rho\sigma}$ and $G^a_{\rho\sigma}$ is the gluon field strength.

\subsection{Interaction with photons only}
\begin{center}
\begin{longtable}{|>{\pnt} r | c | c |}
\hline
n & $D - 4$ & term \\
\hline
& 2 & ${D^\pm}^* D^\pm F^{\mu\nu} F_{\mu\nu}$ \\
\hline
& 2 & ${D^\pm}^* D^\pm F^{\mu\nu} \tilde{F}_{\mu\nu}$ \\

\doubline

& 2 & $i \, ({D^\pm}^* \overleftrightarrow{\covD_\mu} D^\pm) (\partial_\nu F^{\mu\nu})$ \\

\doubline

$\nnn{1}$ & 0 & $(\covD^\mu \covD_\mu {D^\pm}^*) D^\pm + {D^\pm}^* (\covD^\mu \covD_\mu D^\pm)$ \\
\hline
$\nnn{2}$ & 0 & $(\covD^\mu {D^\pm}^*) (\covD_\mu D^\pm)$ \\
\hline
$\nnn{1} + 2 * \nnn{2}$ & 0 & $\partial^\mu \partial_\mu ({D^\pm}^* D^\pm)$ \\
\hline
\multirow{2}{*}{$\nnn{3}$} & \multirow{2}{*}{0} & $i \, [(\covD^\mu \covD_\mu {D^\pm}^*) D^\pm - {D^\pm}^* (\covD^\mu \covD_\mu D^\pm)]$ \\
& & $= - i \, \partial^\mu ({D^\pm}^* \overleftrightarrow{\covD_\mu} D^\pm)$ \\

\doubline

$\nnn{4}$ & 2 &$(\covD^\mu \covD_\mu \covD^\nu \covD_\nu {D^\pm}^*) D^\pm + {D^\pm}^* (\covD^\mu \covD_\mu \covD^\nu \covD_\nu D^\pm)$ \\
\hline
$\nnn{5}$ & 2 &$(\covD^\mu \covD_\mu \covD^\nu {D^\pm}^*) (\covD_\nu D^\pm) + (\covD_\nu {D^\pm}^*) (\covD^\mu \covD_\mu \covD^\nu D^\pm)$ \\
\hline
$\nnn{4i}$ & 2 &$i \, [(\covD^\mu \covD_\mu \covD^\nu \covD_\nu {D^\pm}^*) D^\pm - {D^\pm}^* (\covD^\mu \covD_\mu \covD^\nu \covD_\nu D^\pm)]$ \\
\hline
$\nnn{5i}$ & 2 &$i \, [(\covD^\mu \covD_\mu \covD^\nu {D^\pm}^*) (\covD_\nu D^\pm) - (\covD_\nu {D^\pm}^*) (\covD^\mu \covD_\mu \covD^\nu D^\pm)]$ \\
\hline
$\nnn{6}$ & 2 &$(\covD^\mu \covD_\mu {D^\pm}^*) (\covD^\nu \covD_\nu D^\pm)$ \\
\hline
$\nnn{7}$ & 2 &$(\covD^\mu \covD^\nu {D^\pm}^*) (\covD_\mu \covD_\nu D^\pm)$ \\
\hline

\hline
\end{longtable}
\end{center}
where $\tilde{F}^{\mu\nu} \equiv \varepsilon^{\mu\nu\rho\sigma} F_{\rho\sigma}$ , with $F_{\rho\sigma}$ the photon field strength.

Not all the operators are independent. In fact the following linear combinations amount to total derivatives:
\\
$(\nnn{1} + 2 * \nnn{2})$
\\
$\nnn{3}$
\\
$\nnn{4} + \nnn{5}$
\\
$\nnn{4i} + \nnn{5i}$
\\
$\nnn{5} + \nnn{6}$
\\
$\nnn{5} + \nnn{7}$.

\subsection{Interaction with electroweak gauge bosons $W$, $Z$ and $A$}


\subsubsection{With no derivatives}
\begin{center}
\begin{longtable}{|>{\pnt} r | c | c |}
\hline
n & $D - 4$ & term \\
\hline
& 0 & ${D^0}^* D^\pm W^\mp_\mu Z^\mu + \text{ h.c.}$ \\
\hline
& 0 & $i \, [{D^0}^* D^\pm W^\mp_\mu Z^\mu - \text{ h.c.}]$ \\

\doubline

& 2 & ${D^0}^* D^\pm W^\mp_\mu Z^\mu Z^\nu Z_\nu + \text{ h.c.}$ \\
\hline
& 2 & $i \, [{D^0}^* D^\pm W^\mp_\mu Z^\mu Z^\nu Z_\nu - \text{ h.c.}]$ \\

\doubline

& 2 & ${D^0}^* D^\pm W^\mp_\mu W^{+\mu} W^-_\nu Z^\nu + \text{ h.c.}$ \\
\hline
& 2 & $i \, [{D^0}^* D^\pm W^\mp_\mu W^{+\mu} W^-_\nu Z^\nu - \text{ h.c.}]$ \\

\doubline

& 2 & ${D^0}^* D^\pm W^\mp_\mu W^+_\nu W^{-\mu} Z^\nu + \text{ h.c.}$ \\
\hline
& 2 & $i \, [{D^0}^* D^\pm W^\mp_\mu W^+_\nu W^{-\mu} Z^\nu - \text{ h.c.}]$ \\

\doubline

& 0 & ${D^\pm}^* D^\pm Z^\mu Z_\mu$ \\
\hline
& 0 & ${D^\pm}^* D^\pm W^{+\mu} W^-_\mu$ \\
\hline
& 2 & ${D^\pm}^* D^\pm Z^\mu Z_\mu Z^\nu Z_\nu$ \\
\hline
& 2 & ${D^\pm}^* D^\pm Z^\mu Z_\mu W^{+\nu} W^-_\nu$ \\
\hline
& 2 & ${D^\pm}^* D^\pm Z^\mu Z^\nu W^+_\mu W^-_\nu$ \\
\hline
& 2 & ${D^\pm}^* D^\pm W^{+\mu} W^-_\mu W^{+\nu} W^-_\nu$ \\
\hline
& 2 & ${D^\pm}^* D^\pm W^{+\mu} W^{-\nu} W^+_\mu W^-_\nu$ \\
\hline

\end{longtable}
\end{center}

\subsubsection{With one derivative}\label{D^1 D^2}

\bigskip

To ease the notation we divide the terms in two lists:
\begin{itemize}
\item[i)] the first one contains non-hermitian terms; each non-hermitian term gives rise to two hermitian terms, namely its real and imaginary parts. \item[ii)] in the second list all the terms are already hermitian.
\end{itemize}
\begin{center}
\begin{longtable}{|>{\pnt} r | c | c |}
\hline
n & $D - 4$ & term \\
\hline
& 0 & $(\partial^\mu {D^0}^*) \, D^\pm W^\mp_\mu$ \\
\hline
& 0 & ${D^0}^* (\covD^\mu D^\pm) \, W^\mp_\mu$ \\

\doubline

& 2 & $(\partial^\mu {D^0}^*) \, D^\pm W^\mp_\mu Z^\nu Z_\nu$ \\
\hline
& 2 & ${D^0}^* (\covD^\mu D^\pm) \, W^\mp_\mu Z^\nu Z_\nu$ \\
\hline
& 2 & ${D^0}^* D^\pm W^\mp_\mu \, (\partial^\mu Z^\nu) \, Z_\nu$ \\

\doubline

& 2 & $(\partial_\mu {D^0}^*) \, D^\pm W^\mp_\nu Z^\mu Z^\nu$ \\
\hline
& 2 & ${D^0}^* (\covD_\mu D^\pm) \, W^\mp_\nu Z^\mu Z^\nu$ \\
\hline
& 2 & ${D^0}^* D^\pm W^\mp_\nu \, (\partial_\mu Z^\mu) \, Z^\nu$ \\
\hline
& 2 & ${D^0}^* D^\pm W^\mp_\nu Z^\mu \, (\partial_\mu Z^\nu)$ \\

\doubline

& 2 & $\varepsilon^{\mu\nu\rho\sigma} \, {D^0}^* D^\pm W^\mp_\mu Z_\nu \, (\partial_\rho Z_\sigma)$ \\

\doubline

& 2 & ${D^0}^* D^\pm W^\mp_\mu Z_\nu F^{\mu\nu}$ \\
\hline
& 2 & ${D^0}^* D^\pm W^\mp_\mu Z_\nu \tilde{F}^{\mu\nu}$ \\

\doubline

& 2 & $(\partial^\mu {D^0}^*) D^\pm \, W^\mp_\mu W^{\pm\nu} W^\mp_\nu$ \\
\hline
& 2 & ${D^0}^* (\covD^\mu D^\pm) \, W^\mp_\mu W^{\pm\nu} W^\mp_\nu$ \\
\hline
& 2 & ${D^0}^* D^\pm \, W^\mp_\mu (\covD^\mu W^{\pm\nu}) W^\mp_\nu$ \\
\hline
& 2 & ${D^0}^* D^\pm \, W^\mp_\mu W^{\pm\nu} (\covD^\mu W^\mp_\nu)$ \\

\doubline

& 2 & $(\partial_\mu {D^0}^*) \, D^\pm \, W^{\mp\nu} W^{\pm\mu} W^\mp_\nu$ \\
\hline
& 2 & ${D^0}^* (\covD_\mu D^\pm) \, W^{\mp\nu} W^{\pm\mu} W^\mp_\nu$ \\
\hline
& 2 & ${D^0}^* D^\pm \, (\covD_\mu W^{\mp\nu}) \, W^{\pm\mu} W^\mp_\nu$ \\

\doubline

& 2 & $\varepsilon^{\mu\nu\rho\sigma} \, {D^0}^* D^\pm (\covD_\mu W^\mp_\nu) W^\pm_\rho W^\mp_\sigma$ \\
\hline
\end{longtable}
\end{center}

 We use here the operators defined in \ref{D^1 phi^2}.
\begin{center}
\begin{longtable}{|>{\pnt} r | c | c |}
\hline
n & $D - 4$ & term \\
\hline
& 0 & $\partial_\mu ({D^\pm}^* D^\pm) \, Z^\mu$ \\
\hline
& 0 & $i \, ({D^\pm}^* \overleftrightarrow{\covD_\mu} D^\pm) \, Z^\mu$ \\

\doubline

& 2 & $\partial_\mu ({D^\pm}^* D^\pm) \, Z^\mu Z^\nu Z_\nu$ \\
\hline
& 2 & $i \, ({D^\pm}^* \overleftrightarrow{\covD_\mu} D^\pm) \, Z^\mu Z^\nu Z_\nu$ \\
\hline
& 2 & ${D^\pm}^* D^\pm (\partial_\mu Z^\mu) Z^\nu Z_\nu$ \\

\doubline

& 2 & $\partial_\mu ({D^\pm}^* D^\pm) \, Z^\mu W^{+\nu} W^-_\nu$ \\
\hline
& 2 & $i \, ({D^\pm}^* \overleftrightarrow{\covD_\mu} D^\pm) \, Z^\mu W^{+\nu} W^-_\nu$ \\
\hline
& 2 & ${D^\pm}^* D^\pm (\partial_\mu Z^\mu) W^{+\nu} W^-_\nu$ \\
\hline
& 2 & $i \, {D^\pm}^* D^\pm Z^\mu \, (W^{+\nu} \overleftrightarrow{\covD_\mu} W^-_\nu)$ \\

\doubline

\multirow{2}{*}{$\nnn{1}$} & \multirow{2}{*}{2} & $(\covD_\mu {D^\pm}^*) D^\pm Z^\nu W^{+\mu} W^-_\nu +$ \\
& & ${D^\pm}^* (\covD_\mu D^\pm) Z^\nu W^+_\nu W^{-\mu}$ \\
\hline
\multirow{2}{*}{$\nnn{2}$} & \multirow{2}{*}{2} & ${D^\pm}^* (\covD_\mu D^\pm) Z^\nu W^{+\mu} W^-_\nu +$ \\
& & $(\covD_\mu {D^\pm}^*) D^\pm Z^\nu W^+_\nu W^{-\mu}$ \\
\hline
$\nnn{1} + \nnn{2}$ & 2 & $\partial_\mu ({D^\pm}^* D^\pm) \, Z^\nu \wis{+}$ \\
\hline
$\nnn{2} - \nnn{1}$ & 2 & $({D^\pm}^* \overleftrightarrow{\covD_\mu} D^\pm) \, Z^\nu \wis{-}$ \\
\hline
$\nnn{3}$ & 2 & ${D^\pm}^* D^\pm (\partial_\mu Z^\nu) \, \wis{+}$ \\
\hline
$\nnn{4}$ & 2 & ${D^\pm}^* D^\pm Z^\nu \dwis{1}{+}$ \\
\hline
$\nnn{5}$ & 2 & ${D^\pm}^* D^\pm Z^\nu \dwis{2}{+}$ \\
\hline
$\nnn{4} + \nnn{5}$ & 2 & ${D^\pm}^* D^\pm Z^\nu (\partial_\mu \wis{+})$ \\
\hline
$\nnn{4} - \nnn{5}$ & 2 & ${D^\pm}^* D^\pm Z^\nu (W^-_\nu \overleftrightarrow{\covD_\mu} W^{+\mu} - W^{-\mu} \overleftrightarrow{\covD_\mu} W^+_\nu)$ \\
\hline
\multirow{2}{*}{$\nnn{6}$} & \multirow{2}{*}{2} & $i \, [ (\covD_\mu {D^\pm}^*) D^\pm Z^\nu W^{+\mu} W^-_\nu -$ \\
& & ${D^\pm}^* (\covD_\mu D^\pm) Z^\nu W^+_\nu W^{-\mu} ]$ \\
\hline
\multirow{2}{*}{$\nnn{7}$} & \multirow{2}{*}{2} & $i [ {D^\pm}^* (\covD_\mu D^\pm) Z^\nu W^{+\mu} W^-_\nu -$ \\
& & $(\covD_\mu {D^\pm}^*) D^\pm Z^\nu W^+_\nu W^{-\mu} ]$ \\
\hline
$\nnn{6} + \nnn{7}$ & 2 & $i \, \partial_\mu ({D^\pm}^* D^\pm) \, Z^\nu \wis{-}$ \\
\hline
$\nnn{7} - \nnn{6}$ & 2 & $i \, ({D^\pm}^* \overleftrightarrow{\covD_\mu} D^\pm) Z^\nu \wis{+}$ \\
\hline
$\nnn{8}$ & 2 & $i \, {D^\pm}^* D^\pm (\partial_\mu Z^\nu) \, \wis{-}$ \\
\hline
$\nnn{9}$ & 2 & $i \, {D^\pm}^* D^\pm Z^\nu \dwis{1}{-}$ \\
\hline
$\nnn{10}$ & 2 & $i \, {D^\pm}^* D^\pm Z^\nu \dwis{2}{-}$ \\
\hline
$\nnn{9} + \nnn{10}$ & 2 & $i \, {D^\pm}^* D^\pm Z^\nu (\partial_\mu \wis{-})$ \\
\hline
$\nnn{9} - \nnn{10}$ & 2 & $i \, {D^\pm}^* D^\pm Z^\nu (W^-_\nu \overleftrightarrow{\covD_\mu} W^{+\mu} + W^{-\mu} \overleftrightarrow{\covD_\mu} W^+_\nu)$ \\

\doubline

& 2 & $\varepsilon^{\mu\nu\rho\sigma} ({D^\pm}^* \overleftrightarrow{\covD_\mu} D^\pm) \, Z_\nu W^+_\rho W^-_\sigma$ \\
\hline
$\nnn{11}$ & 2 & $i \, \varepsilon^{\mu\nu\rho\sigma} \partial_\mu ({D^\pm}^* D^\pm) \, Z_\nu W^+_\rho W^-_\sigma$ \\
\hline
$\nnn{12}$ & 2 & $i \, \varepsilon^{\mu\nu\rho\sigma} {D^\pm}^* D^\pm (\partial_\mu Z_\nu) W^+_\rho W^-_\sigma$ \\
\hline
$\nnn{13}$ & 2 & $i \, \varepsilon^{\mu\nu\rho\sigma} {D^\pm}^* D^\pm Z_\nu \, \partial_\mu (W^+_\rho W^-_\sigma)$ \\
\hline
& 2 & $\varepsilon^{\mu\nu\rho\sigma} {D^\pm}^* D^\pm Z_\nu \, (W^+_\rho \overleftrightarrow{\covD_\mu} W^-_\sigma)$ \\

\doubline

& 2 & $i \, {D^\pm}^* D^\pm W^{+\mu} W^{-\nu} F_{\mu\nu}$ \\
\hline
& 2 & $i \, {D^\pm}^* D^\pm W^{+\mu} W^{-\nu} \tilde{F}_{\mu\nu}$ \\

\hline
\end{longtable}
\end{center}
The following combinations are total derivatives:
\\
$(\nnn{1} + \nnn{2}) + \nnn{3} + (\nnn{4} + \nnn{5})$
\\
$(\nnn{6} + \nnn{7}) + \nnn{8} + (\nnn{9} + \nnn{10})$
\\
$\nnn{11} + \nnn{12} + \nnn{13}$.

\subsubsection{With two derivatives}\label{D^2 D^2}
For simplicity and readability we list here only some terms, the rest being derived from these by permitting the derivates to act on the various fields in different order. When below a term the symbol \virg{$*$} appears this means that the remaining terms obtained by simply changing position of the derivatives are not displayed here but should be taken into account. One has also to pay attention to the fact that a partial derivative turns into a covariant one when it acts on a charged field. 
\begin{center}
\begin{longtable}{|>{\pnt} r | c | c |}
\hline
n & $D - 4$ & term \\
\hline
& 2 & $(\partial^\mu \partial_\mu {D^0}^*) \, D^\pm W^\mp_\nu Z^\nu + \text{ h.c.}$ \\
\hline
& 2 & * \\

\doubline

& 2 & $i \, [(\partial^\mu \partial_\mu {D^0}^*) \, D^\pm W^\mp_\nu Z^\nu - \text{ h.c.}]$ \\
\hline
& 2 & * \\

\doubline

& 2 & $(\partial^\mu \partial_\nu {D^0}^*) \, D^\pm W^\mp_\mu Z^\nu + \text{ h.c.}$ \\
\hline
& 2 & * \\

\doubline

& 2 & $i \, [(\partial^\mu \partial_\nu {D^0}^*) \, D^\pm W^\mp_\mu Z^\nu - \text{ h.c.}]$ \\
\hline
& 2 & * \\

\doubline

& 2 & $\varepsilon^{\mu\nu\rho\sigma} \, (\partial_\mu {D^0}^*) \, (\covD_\nu D^\pm) \, W^\mp_\rho Z_\sigma + \text{ h.c.}$ \\
\hline
& 2 & * \\

\doubline

& 2 & $i \, [\varepsilon^{\mu\nu\rho\sigma} \, (\partial_\mu {D^0}^*) \, (\covD_\nu D^\pm) \, W^\mp_\rho Z_\sigma - \text{ h.c.}]$ \\
\hline
& 2 & * \\

\doubline

& 2 & $(\partial_\mu {D^0}^*) \, D^\pm W^\mp_\nu F^{\mu\nu} + \text{ h.c.}$ \\
\hline
& 2 & * \\

\doubline

& 2 & $i \, [(\partial_\mu {D^0}^*) \, D^\pm W^\mp_\nu F^{\mu\nu} - \text{ h.c.}]$ \\
\hline
& 2 & * \\

\doubline

& 2 & $(\partial_\mu {D^0}^*) \, D^\pm W^\mp_\nu \tilde{F}^{\mu\nu} + \text{ h.c.}$ \\
\hline
& 2 & * \\

\doubline

& 2 & $i \, [(\partial_\mu {D^0}^*) \, D^\pm W^\mp_\nu \tilde{F}^{\mu\nu} - \text{ h.c.}]$ \\
\hline
& 2 & * \\

\doubline

$\nnn{1}$ & 2 & $[(\covD^\mu \covD_\mu {D^\pm}^*) \, D^\pm + {D^\pm}^* \, (\covD^\mu \covD_\mu D^\pm)] \, Z^\nu Z_\nu$ \\
\hline
$\nnn{2}$ & 2 & $(\covD^\mu {D^\pm}^*) \, (\covD_\mu D^\pm) \, Z^\nu Z_\nu$ \\
\hline
$\nnn{1} + 2 * \nnn{2}$ & 2 & $\partial^\mu \partial_\mu ({D^\pm}^* D^\pm) \, Z^\nu Z_\nu$ \\
\hline
& \multirow{2}{*}{2} & $i \, [(\covD^\mu \covD_\mu {D^\pm}^*) \, D^\pm - {D^\pm}^* \,(\covD^\mu \covD_\mu D^\pm)] \, Z^\nu Z_\nu$ \\
& & $= - i \, \partial^\mu ({D^\pm}^* \overleftrightarrow{\covD_\mu} D^\pm) \, Z^\nu Z_\nu$ \\
\hline
& 2 & ${D^\pm}^* D^\pm \, (\partial^\mu Z^\nu) \, (\partial_\mu Z_\nu)$ \\

\doubline

$\nnn{3}$ & 2 & $[(\covD_\mu \covD_\nu {D^\pm}^*) \, D^\pm + {D^\pm}^* \, (\covD_\mu \covD_\nu D^\pm)] \, Z^\mu Z^\nu$ \\
\hline
$\nnn{4}$ & 2 & $(\covD_\mu {D^\pm}^*) \, (\covD_\nu D^\pm) \, Z^\mu Z^\nu$ \\
\hline
$\nnn{3} + 2 * \nnn{4}$ & 2 & $\partial_\mu \partial_\nu ({D^\pm}^* D^\pm) \, Z^\mu Z^\nu$ \\
\hline
& \multirow{2}{*}{2} & $i \, [(\covD_\mu \covD_\nu {D^\pm}^*) \, D^\pm - {D^\pm}^* \,(\covD_\mu \covD_\nu D^\pm)] \, Z^\mu Z^\nu$ \\
& & $= - i \, \partial_\mu ({D^\pm}^* \overleftrightarrow{\covD_\nu} D^\pm) \, Z^\mu Z^\nu$ \\
\hline
$\nnn{31}$ & 2 & $\partial_\mu ({D^\pm}^* D^\pm) \, (\partial_\nu Z^\mu) Z^\nu$ \\
\hline
$\nnn{32}$ & 2 & $i \, ({D^\pm}^* \overleftrightarrow{\covD_\mu} D^\pm) \, (\partial_\nu Z^\mu) Z^\nu$ \\
\hline
$\nnn{33}$ & 2 & $\partial_\mu ({D^\pm}^* D^\pm) \, Z^\mu (\partial_\nu Z^\nu)$ \\
\hline
$\nnn{34}$ & 2 & $i \, ({D^\pm}^* \overleftrightarrow{\covD_\mu} D^\pm) \, Z^\mu (\partial_\nu Z^\nu)$ \\
\hline
$\nnn{33} + \nnn{31}$ & 2 & $\partial_\mu ({D^\pm}^* D^\pm) \, \partial_\nu (Z^\mu Z^\nu)$ \\
\hline
$\nnn{33} - \nnn{31}$ & 2 & $\partial_\mu ({D^\pm}^* D^\pm) \, (Z^\mu \overleftrightarrow{\partial_\nu} Z^\nu)$ \\
\hline
$\nnn{34} + \nnn{32}$ & 2 & $i \, ({D^\pm}^* \overleftrightarrow{\covD_\mu} D^\pm) \, \partial_\nu (Z^\mu Z^\nu)$ \\
\hline
$\nnn{34} - \nnn{32}$ & 2 & $i \, ({D^\pm}^* \overleftrightarrow{\covD_\mu} D^\pm) \, (Z^\mu \overleftrightarrow{\partial_\nu} Z^\nu)$ \\
\hline
$\nnn{35}$ & 2 & ${D^\pm}^* D^\pm \, (\partial_\mu Z^\mu) \, (\partial_\nu Z^\nu)$ \\
\hline
$\nnn{36}$ & 2 & ${D^\pm}^* D^\pm \, (\partial_\mu Z^\nu) \, (\partial_\nu Z^\mu)$ \\
\hline
$\nnn{37}$ & 2 & ${D^\pm}^* D^\pm \, (\partial_\mu \partial_\nu Z^\mu) \, Z^\nu$ \\
\hline
$\nnn{35} + \nnn{36}$ & \multirow{2}{*}{2} & \multirow{2}{*}{${D^\pm}^* D^\pm \, \partial_\mu \partial_\nu (Z^\mu Z^\nu)$} \\
$+ 2 * \nnn{37}$ & & \\
\hline
$\nnn{35} - \nnn{36}$ & 2 & ${D^\pm}^* D^\pm \, \partial_\mu (Z^\mu \overleftrightarrow{\partial_\nu} Z^\nu)$ \\

\doubline

$\nnn{38}$ & 2 & $\varepsilon^{\mu\nu\rho\sigma} \partial_\mu ({D^\pm}^* D^\pm) \, (Z_\rho \overleftrightarrow{\partial_\nu} Z_\sigma)$ \\
\hline
& 2 & $i \, \varepsilon^{\mu\nu\rho\sigma} ({D^\pm}^* \overleftrightarrow{\covD_\mu} D^\pm) \, (Z_\rho \overleftrightarrow{\partial_\nu} Z_\sigma)$ \\
\hline
$\nnn{39}$ & 2 & $\varepsilon^{\mu\nu\rho\sigma} {D^\pm}^* D^\pm \, (\partial_\mu Z_\rho) (\partial_\nu Z_\sigma)$ \\

\doubline

$\nnn{5}$ & 2 & $[(\covD^\mu \covD_\mu {D^\pm}^*) \, D^\pm + {D^\pm}^* \, (\covD^\mu \covD_\mu D^\pm)] \, W^{+\nu} W^-_\nu$ \\
\hline
$\nnn{6}$ & 2 & $(\covD^\mu {D^\pm}^*) \, (\covD_\mu D^\pm) \, W^{+\nu} W^-_\nu$ \\
\hline
$\nnn{5} + 2 * \nnn{6}$ & 2 & $\partial^\mu \partial_\mu ({D^\pm}^* D^\pm) \, W^{+\nu} W^-_\nu$ \\
\hline
\multirow{2}{*}{$\nnn{40}$} & \multirow{2}{*}{2} & $i \, [(\covD^\mu \covD_\mu {D^\pm}^*) \, D^\pm - {D^\pm}^* \,(\covD^\mu \covD_\mu D^\pm)] \, W^{+\nu} W^-_\nu$ \\
& & $= - i \, \partial^\mu ({D^\pm}^* \overleftrightarrow{\covD_\mu} D^\pm) \, W^{+\nu} W^-_\nu$ \\
\hline
$\nnn{41}$ & 2 & $\partial^\mu ({D^\pm}^* D^\pm) \, \partial_\mu (W^{+\nu} W^-_\nu)$ \\
\hline
$\nnn{42}$ & 2 & $i \, \partial^\mu ({D^\pm}^* D^\pm) \, (W^{+\nu} \overleftrightarrow{\covD_\mu} W^-_\nu)$ \\
\hline
$\nnn{43}$ & 2 & $i \, ({D^\pm}^* \overleftrightarrow{\covD^\mu} D^\pm) \, \partial_\mu (W^{+\nu} W^-_\nu)$ \\
\hline
& 2 & $({D^\pm}^* \overleftrightarrow{\covD^\mu} D^\pm) \, (W^{+\nu} \overleftrightarrow{\covD_\mu} W^-_\nu)$ \\

\doubline

$\nnn{7}$ & 2 & $[(\covD_\mu \covD_\nu {D^\pm}^*) \, D^\pm + {D^\pm}^* \, (\covD_\mu \covD_\nu D^\pm)] \, W^{+\mu} W^{-\nu}$ \\
\hline
$\nnn{8a}$ & 2 & $(\covD_\mu {D^\pm}^*) \, (\covD_\nu D^\pm) \, W^{+\mu} W^{-\nu}$ \\
\hline
$\nnn{8b}$ & 2 & $(\covD_\mu {D^\pm}^*) \, (\covD_\nu D^\pm) \, W^{+\nu} W^{-\mu}$ \\
\hline
$\nnn{7} + \nnn{8a} + \nnn{8b}$ & 2 & $\partial_\mu \partial_\nu ({D^\pm}^* D^\pm) \, W^{+\mu} W^{-\nu}$ \\
\hline
$(\nnn{8a} - \nnn{8b}) / 2$ & 2 & $\partial_\mu ({D^\pm}^* \overleftrightarrow{\covD^\nu} D^\pm) \, \wis{-}$ \\
\hline
\multirow{2}{*}{$\nnn{7i}$} & \multirow{2}{*}{2} & $i \, \partial_\mu ({D^\pm}^* \overleftrightarrow{\covD^\nu} D^\pm) \, \wis{+} =$ \\
& & $2 \, i \, [(\covD_\mu \covD_\nu {D^\pm}^*) \, D^\pm - {D^\pm}^* \, (\covD_\mu \covD_\nu D^\pm)] \, W^{+\mu} W^{-\nu}$ \\
\hline

\multirow{2}{*}{$\nnn{51}$} & \multirow{2}{*}{2} & $(\covD_\mu {D^\pm}^*) D^\pm \, W^{+\mu} (\covD_\nu W^{-\nu})$ \\
& & $+ \, {D^\pm}^* (\covD_\mu D^\pm) \, (\covD_\nu W^{+\nu}) W^{-\mu}$ \\
\hline
\multirow{2}{*}{$\nnn{51i}$} & \multirow{2}{*}{2} & $i \, [(\covD_\mu {D^\pm}^*) D^\pm \, W^{+\mu} (\covD_\nu W^{-\nu})$ \\
& & $- \, {D^\pm}^* (\covD_\mu D^\pm) \, (\covD_\nu W^{+\nu}) W^{-\mu}]$ \\
\hline
\multirow{2}{*}{$\nnn{52}$} & \multirow{2}{*}{2} & $(\covD_\mu {D^\pm}^*) D^\pm \, (\covD_\nu W^{+\nu}) W^{-\mu}$ \\
& & $+ \, {D^\pm}^* (\covD_\mu D^\pm) \, W^{+\mu} (\covD_\nu W^{-\nu})$ \\
\hline
\multirow{2}{*}{$\nnn{52i}$} & \multirow{2}{*}{2} & $i \, [(\covD_\mu {D^\pm}^*) D^\pm \, (\covD_\nu W^{+\nu}) W^{-\mu}$ \\
& & $- \, {D^\pm}^* (\covD_\mu D^\pm) \, W^{+\mu} (\covD_\nu W^{-\nu})]$ \\
\hline

\multirow{2}{*}{$\nnn{53}$} & \multirow{2}{*}{2} & $(\covD_\mu {D^\pm}^*) D^\pm \, (\covD_\nu W^{+\mu}) W^{-\nu}$ \\
& & $+ \, {D^\pm}^* (\covD_\mu D^\pm) \, W^{+\nu} (\covD_\nu W^{-\mu})$ \\
\hline
\multirow{2}{*}{$\nnn{53i}$} & \multirow{2}{*}{2} & $i \, [(\covD_\mu {D^\pm}^*) D^\pm \, (\covD_\nu W^{+\mu}) W^{-\nu}$ \\
& & $- \, {D^\pm}^* (\covD_\mu D^\pm) \, W^{+\nu} (\covD_\nu W^{-\mu})]$ \\
\hline
\multirow{2}{*}{$\nnn{54}$} & \multirow{2}{*}{2} & $(\covD_\mu {D^\pm}^*) D^\pm \, W^{+\nu} (\covD_\nu W^{-\mu})$ \\
& & $+ \, {D^\pm}^* (\covD_\mu D^\pm) \, (\covD_\nu W^{+\mu}) W^{-\nu}$ \\
\hline
\multirow{2}{*}{$\nnn{54i}$} & \multirow{2}{*}{2} & $i \, [(\covD_\mu {D^\pm}^*) D^\pm \, W^{+\nu} (\covD_\nu W^{-\mu})$ \\
& & $- \, {D^\pm}^* (\covD_\mu D^\pm) \, (\covD_\nu W^{+\mu}) W^{-\nu}]$ \\
\hline
$\nnn{51} + \nnn{52}$ & 2 & $\partial^\nu ({D^\pm}^* D^\pm) \, \dwis{1}{+}$ \\
\hline
$\nnn{51} - \nnn{52}$ & 2 & $({D^\pm}^* \overleftrightarrow{\covD^\nu} D^\pm) \, \dwis{1}{-}$ \\
\hline
$- (\nnn{51i} + \nnn{52i})$ & 2 & $i \, ({D^\pm}^* \overleftrightarrow{\covD^\nu} D^\pm) \, \dwis{1}{+}$ \\
\hline
$\nnn{52i} - \nnn{51i}$ & 2 & $i \, \partial^\nu ({D^\pm}^* D^\pm) \, \dwis{1}{-}$ \\
\hline
$\nnn{53} + \nnn{54}$ & 2 & $\partial^\nu ({D^\pm}^* D^\pm) \, \dwis{2}{+}$ \\
\hline
$\nnn{53} - \nnn{54}$ & 2 & $({D^\pm}^* \overleftrightarrow{\covD^\nu} D^\pm) \, \dwis{2}{-}$ \\
\hline
$- (\nnn{53i} + \nnn{54i})$ & 2 & $i \, ({D^\pm}^* \overleftrightarrow{\covD^\nu} D^\pm) \, \dwis{2}{+}$ \\
\hline
$\nnn{54i} - \nnn{53i}$ & 2 & $i \, \partial^\nu ({D^\pm}^* D^\pm) \, \dwis{2}{-}$ \\
\hline
$\nnn{61}$ & 2 & ${D^\pm}^* D^\pm \, (\covD_\mu W^{+\mu}) \, (\covD_\nu W^{-\nu})$ \\
\hline
$\nnn{62}$ & 2 & ${D^\pm}^* D^\pm \, (\covD_\mu W^{+\nu}) \, (\covD_\nu W^{-\mu})$ \\
\hline
$\nnn{63}$ & 2 & ${D^\pm}^* D^\pm \, (\covD^\mu W^{+\nu}) \, (\covD_\mu W^-_\nu)$ \\
\hline
$\nnn{64a}$ & 2 & ${D^\pm}^* D^\pm \, [(\covD_\mu \covD_\nu W^{+\mu}) \, W^{-\nu} + W^{+\mu} \, (\covD_\mu \covD_\nu W^{-\nu})]$ \\
\hline
\multirow{3}{*}{$\nnn{64b}$} & \multirow{3}{*}{2} & $2 \, i \, {D^\pm}^* D^\pm \, [(\covD_\mu \covD_\nu W^{+\mu}) \, W^{-\nu} - W^{+\mu} \, (\covD_\mu \covD_\nu W^{-\nu})]$ \\
& & $= i \, {D^\pm}^* D^\pm \, \partial_\mu (W^{-\nu} \overleftrightarrow{\covD_\nu} W^{+\mu} + W^{-\mu} \overleftrightarrow{\covD_\nu} W^{+\nu})$ \\
& & $= i \, {D^\pm}^* D^\pm \, \partial^\nu (\dwis{2}{-} - \dwis{1}{-})$ \\
\hline
$\nnn{65a}$ & 2 & ${D^\pm}^* D^\pm \, [(\covD^\mu \covD_\mu W^{+\nu}) \, W^-_\nu + W^{+\nu} \, (\covD^\mu \covD_\mu W^-_\nu)]$ \\
\hline
\multirow{2}{*}{$\nnn{65b}$} & \multirow{2}{*}{2} & $i \, {D^\pm}^* D^\pm \, [(\covD^\mu \covD_\mu W^{+\nu}) \, W^-_\nu - W^{+\nu} \, (\covD^\mu \covD_\mu W^-_\nu)]$ \\
& & $= i \, {D^\pm}^* D^\pm \, \partial^\mu (W^{-\nu} \overleftrightarrow{\covD_\mu} W^+_\nu)$ \\
\hline
$2 * \nnn{63} + \nnn{65a}$ & 2 & ${D^\pm}^* D^\pm \, \partial^\mu \partial_\mu (W^{+\nu} W^-_\nu)$ \\
\hline
$\nnn{61} + \nnn{62} + \nnn{64a}$ & 2 & ${D^\pm}^* D^\pm \, \partial_\mu \partial_\nu (W^{+\mu} W^{-\nu})$ \\
\hline
\multirow{2}{*}{$2 * (\nnn{62} - \nnn{61})$} & \multirow{2}{*}{2} & ${D^\pm}^* D^\pm \, \partial_\mu (W^{-\nu} \overleftrightarrow{\covD_\nu} W^{+\mu} - W^{-\mu} \overleftrightarrow{\covD_\nu} W^{+\nu})$ \\
& & $= {D^\pm}^* D^\pm \, \partial^\nu (\dwis{2}{+} - \dwis{1}{+})$ \\

\doubline

$\nnn{9}$ & 2 & $\varepsilon^{\mu\nu\rho\sigma} (\covD_\mu {D^\pm}^*) (\covD_\nu D^\pm) \, W^+_\rho W^-_\sigma$ \\
\hline
\multirow{2}{*}{$\nnn{10a}$} & \multirow{2}{*}{2} & $\varepsilon^{\mu\nu\rho\sigma} [(\covD_\mu {D^\pm}^*) \, D^\pm \, (\covD_\nu W^+_\rho) \, W^-_\sigma$ \\
& & $- {D^\pm}^* \, (\covD_\mu D^\pm) \, W^+_\rho \, (\covD_\nu W^-_\sigma)]$ \\
\hline
\multirow{2}{*}{$\nnn{10b}$} & \multirow{2}{*}{2} & $i \, \varepsilon^{\mu\nu\rho\sigma} [(\covD_\mu {D^\pm}^*) \, D^\pm \, (\covD_\nu W^+_\rho) \, W^-_\sigma$ \\
& & $+ {D^\pm}^* \, (\covD_\mu D^\pm) \, W^+_\rho \, (\covD_\nu W^-_\sigma)]$ \\
\hline
\multirow{2}{*}{$\nnn{11a}$} & \multirow{2}{*}{2} & $\varepsilon^{\mu\nu\rho\sigma} [(\covD_\mu {D^\pm}^*) \, D^\pm \, W^+_\rho \, (\covD_\nu W^-_\sigma)$ \\
& & $- {D^\pm}^* \, (\covD_\mu D^\pm) \, (\covD_\nu W^+_\rho) \, W^-_\sigma]$ \\
\hline
\multirow{2}{*}{$\nnn{11b}$} & \multirow{2}{*}{2} & $i \, \varepsilon^{\mu\nu\rho\sigma} [(\covD_\mu {D^\pm}^*) \, D^\pm \, W^+_\rho \, (\covD_\nu W^-_\sigma)$ \\
& & $+ {D^\pm}^* \, (\covD_\mu D^\pm) \, (\covD_\nu W^+_\rho) \, W^-_\sigma]$ \\
\hline
$- (\nnn{10a} + \nnn{11a})$ & 2 & $\varepsilon^{\mu\nu\rho\sigma} ({D^\pm}^* \overleftrightarrow{\covD_\mu} D^\pm) \, \partial_\nu (W^+_\rho W^-_\sigma)$ \\
\hline
$\nnn{11a} - \nnn{10a}$ & 2 & $\varepsilon^{\mu\nu\rho\sigma} \partial_\mu ({D^\pm}^* D^\pm) \, (W^+_\rho \overleftrightarrow{\covD_\nu} W^-_\sigma)$ \\
\hline
$\nnn{10b} + \nnn{11b}$ & 2 & $i \, \varepsilon^{\mu\nu\rho\sigma} \partial_\mu ({D^\pm}^* D^\pm) \, \partial_\nu (W^+_\rho W^-_\sigma) \xrightarrow{\int \text{by parts}} 0$ \\
\hline
$\nnn{10b} - \nnn{11b}$ & 2 & $i \, \varepsilon^{\mu\nu\rho\sigma} ({D^\pm}^* \overleftrightarrow{\covD_\mu} D^\pm) \, (W^+_\rho \overleftrightarrow{\covD_\nu} W^-_\sigma)$ \\
\hline
$\nnn{12}$ & 2 & $\varepsilon^{\mu\nu\rho\sigma} {D^\pm}^* D^\pm \, (\covD_\mu W^+_\rho) (\covD_\nu W^-_\sigma)$ \\

\doubline

& 2 & $\partial_\mu ({D^\pm}^* D^\pm) \, Z_\nu F^{\mu\nu}$ \\
\hline
& 2 & ${D^\pm}^* D^\pm \, (\partial_\mu Z_\nu) F^{\mu\nu}$ \\
\hline
& 2 & $i \, ({D^\pm}^* \overleftrightarrow{\covD_\mu} D^\pm) \, Z_\nu F^{\mu\nu}$ \\

\doubline

& 2 & $\partial_\mu ({D^\pm}^* D^\pm) \, Z_\nu \tilde{F}^{\mu\nu}$ \\
\hline
& 2 & $i \, ({D^\pm}^* \overleftrightarrow{\covD_\mu} D^\pm) \, Z_\nu \tilde{F}^{\mu\nu}$ \\

\hline
\end{longtable}
\end{center}


The following combinations are total derivatives:
\\
$2 * \nnn{9} + \nnn{10a} + \nnn{11a}$ and $\nnn{11a} - \nnn{10a} + 2 * \nnn{12}$
\\
\\
$(\nnn{3} + 2 * \nnn{4}) + (\nnn{33} + \nnn{31})$ and $(\nnn{33} + \nnn{31}) + (\nnn{35} + \nnn{36} + 2 * \nnn{37})$
\\
These terms enjoy the relation:
\\
$(\nnn{3} + 2 * \nnn{4}) + 2 * (\nnn{33} + \nnn{31}) + (\nnn{35} + \nnn{36} + 2 * \nnn{37}) = \partial_\mu \partial_\nu ({D^\pm}^* D^\pm Z^\mu Z^\nu)$
\\
\\
$(\nnn{33} - \nnn{31}) + (\nnn{35} - \nnn{36})$
\\
\\
$2 * \nnn{39} + \nnn{38}$
\\
\\
$(\nnn{5} + 2 * \nnn{6}) + \nnn{41}$ and $\nnn{41} + (2 * \nnn{63} + \nnn{65a})$
\\
These terms enjoy the relation:
\\
$(\nnn{5} + 2 * \nnn{6}) + 2 * \nnn{41} + (2 * \nnn{63} + \nnn{65a}) = \partial^\mu \partial_\mu ({D^\pm}^* D^\pm \, W^{+\nu} W^-_\nu)$
\\
\\
$\nnn{42} - \nnn{65b}$
\\
\\
$\nnn{43} - \nnn{40}$
\\
\\
$2 * (\nnn{7} + \nnn{8a} + \nnn{8b}) + (\nnn{51} + \nnn{52}) + (\nnn{53} + \nnn{54})$ and $(\nnn{51} + \nnn{52}) + (\nnn{53} + \nnn{54}) + 2 * (\nnn{61} + \nnn{62} + \nnn{64a})$
\\
These terms enjoy the relation:
\\
$(\nnn{7} + \nnn{8a} + \nnn{8b}) + (\nnn{51} + \nnn{52}) + (\nnn{53} + \nnn{54}) + (\nnn{61} + \nnn{62} + \nnn{64a}) = \partial_\mu \partial_\nu ({D^\pm}^* D^\pm \, W^{+\mu} W^-_\nu)$
\\
\\
$(\nnn{51} - \nnn{52}) + (\nnn{53} - \nnn{54}) + (\nnn{8a} - \nnn{8b})/2$
\\
\\
$- (\nnn{51i} + \nnn{52i}) - (\nnn{53i} + \nnn{54i}) + \nnn{7i}$
\\
\\
$(\nnn{51} + \nnn{52}) - (\nnn{53} + \nnn{54}) - 2 * (\nnn{62} - \nnn{61})$ and $(\nnn{52i} - \nnn{51i}) - (\nnn{54i} - \nnn{53i}) - \nnn{64b}$.

\subsubsection{With three derivatives}

Only one weak boson field at a time can enter the terms with three derivatives. The interaction terms involving the photon and the $W$ are
\begin{center}
\begin{longtable}{|>{\pnt} r | c | c |}
\hline
n & $D - 4$ & term \\
\hline
& 2 & $(\partial^\mu \partial_\mu \partial^\nu {D^0}^*) \, D^\pm W^\mp_\nu + \text{ h.c.}$ \\
\hline
& 2 & * \\

\doubline

& 2 & $i \, [(\partial^\mu \partial_\mu \partial^\nu {D^0}^*) \, D^\pm W^\mp_\nu - \text{ h.c.}]$ \\
\hline
& 2 & * \\

\doubline

& 2 & $\varepsilon^{\mu\nu\rho\sigma} \, (\partial_\mu {D^0}^*) \, (\covD_\nu D^\pm) \, (\covD_\rho W^\mp_\sigma) + \text{ h.c.} \xrightarrow{\int \text{by parts}} 0$ \\

\doubline

& 2 & $i \, \varepsilon^{\mu\nu\rho\sigma} \, [(\partial_\mu {D^0}^*) \, (\covD_\nu D^\pm) \, (\covD_\rho W^\mp_\sigma) - \text{ h.c.}] \xrightarrow{\int \text{by parts}} 0$ \\

\hline
\end{longtable}
\end{center}

The interaction terms involving the photon and the $Z$ are instead
\begin{center}
\begin{longtable}{|>{\pnt} r | c | c |}
\hline
n & $D - 4$ & term \\
\hline
$\nnn{1}$ & 2 & $ {D^\pm}^* D^\pm \, (\partial^2 \partial_\mu Z^\mu)$ \\

\doubline

$\nnn{2}$ & 2 & $\partial_\mu ({D^\pm}^* D^\pm) \, (\partial^2 Z^\mu)$ \\
\hline
$\nnn{3}$ & 2 & $\partial^\mu ({D^\pm}^* D^\pm) \, (\partial_\mu \partial_\nu Z^\nu)$ \\
\hline
$\nnn{4}$ & 2 & $i \, ({D^\pm}^* \overleftrightarrow{\covD_\mu} D^\pm) \, (\partial^2 Z^\mu)$ \\
\hline
$\nnn{5}$ & 2 & $i \, ({D^\pm}^* \overleftrightarrow{\covD^\mu} D^\pm) \, (\partial_\mu \partial_\nu Z^\nu)$ \\

\doubline

$\nnn{6}$ & 2 & $(\covD^\mu {D^\pm}^*) (\covD_\mu D^\pm) \, (\partial_\nu Z^\nu)$ \\
\hline
$\nnn{7}$ & 2 & $[(\covD_\mu {D^\pm}^*) (\covD_\nu D^\pm) + (\covD_\nu {D^\pm}^*) (\covD_\mu D^\pm)] \, (\partial^\mu Z^\nu)$ \\
\hline
$\nnn{8}$ & 2 & $i \, [(\covD_\mu {D^\pm}^*) (\covD_\nu D^\pm) - (\covD_\nu {D^\pm}^*) (\covD_\mu D^\pm)] \, (\partial^\mu Z^\nu)$ \\
\hline
$\nnn{9}$ & 2 & $i \, \varepsilon^{\mu\nu\rho\sigma} \, (\covD_\mu {D^\pm}^*) (\covD_\nu D^\pm) \, (\partial_\rho Z_\sigma) \xrightarrow{\int \text{by parts}} 0$ \\

\doubline

$\nnn{10}$ & 2 & $[(D^2 {D^\pm}^*) D^\pm + {D^\pm}^* (D^2 D^\pm)] \, (\partial_\mu Z^\mu)$ \\
\hline
\multirow{2}{*}{$\nnn{11}$} & \multirow{2}{*}{2} & $- i \, [(D^2 {D^\pm}^*) D^\pm - {D^\pm}^* (D^2 D^\pm)] \, (\partial_\mu Z^\mu)$ \\
& & $= i \, \partial_\mu ({D^\pm}^* \overleftrightarrow{\covD^\mu} D^\pm) \, (\partial_\nu Z^\nu)$ \\
\hline
$\nnn{12}$ & 2 & $[(\covD_\mu \covD_\nu {D^\pm}^*) D^\pm + {D^\pm}^* (\covD_\mu \covD_\nu D^\pm)] \, (\partial^\mu Z^\nu)$ \\
\hline
$\nnn{13}$ & 2 & $i \, [(\covD_\mu \covD_\nu {D^\pm}^*) D^\pm - {D^\pm}^* (\covD_\mu \covD_\nu D^\pm)] \, (\partial^\mu Z^\nu)$ \\
\hline
$2 * \nnn{6} + \nnn{10}$ & 2 & $\partial^\mu \partial_\mu ({D^\pm}^* D^\pm) \, (\partial_\nu Z^\nu)$ \\
\hline
$\nnn{7} + \nnn{12}$ & 2 & $\partial_\mu \partial_\nu ({D^\pm}^* D^\pm) \, (\partial^\mu Z^\nu)$ \\
\hline
$\nnn{8} - \nnn{13}$ & 2 & $i \, \partial^\mu ({D^\pm}^* \overleftrightarrow{\covD_\nu} D^\pm) \, (\partial_\mu Z^\nu)$ \\
\hline
$- (\nnn{8} + \nnn{13})$ & 2 & $i \, \partial_\mu ({D^\pm}^* \overleftrightarrow{\covD^\nu} D^\pm) \, (\partial_\nu Z^\mu)$ \\

\doubline

$\nnn{14}$ & 2 & $[(D^2 {D^\pm}^*) (\covD_\mu D^\pm) + (\covD_\mu {D^\pm}^*) (D^2 D^\pm)] \, Z^\mu$ \\
\hline
$\nnn{15}$ & 2 & $i \, [(D^2 {D^\pm}^*) (\covD_\mu D^\pm) - (\covD_\mu {D^\pm}^*) (D^2 D^\pm)] \, Z^\mu$ \\
\hline
\multirow{2}{*}{$\nnn{16}$} & \multirow{2}{*}{2} & $[(\covD_\mu \covD_\nu {D^\pm}^*) (\covD^\mu D^\pm) + (\covD^\mu {D^\pm}^*) (\covD_\mu \covD_\nu D^\pm)] \, Z^\nu$ \\
& & $= \partial_\mu [(\covD^\nu {D^\pm}^*) (\covD_\nu D^\pm)] \, Z^\mu$ \\
\hline
$\nnn{17}$ & 2 & $i \, [(\covD_\mu \covD_\nu {D^\pm}^*) (\covD^\mu D^\pm) - (\covD^\mu {D^\pm}^*) (\covD_\mu \covD_\nu D^\pm)] \, Z^\nu$ \\
\hline
$\nnn{18}$ & 2 & $[(D^2 \covD_\mu {D^\pm}^*) D^\pm + {D^\pm}^* (D^2 \covD_\mu D^\pm)] \, Z^\mu$ \\
\hline
$\nnn{19}$ & 2 & $i \, [(D^2 \covD_\mu {D^\pm}^*) D^\pm - {D^\pm}^* (D^2 \covD_\mu D^\pm)] \, Z^\mu$ \\
\hline
$\nnn{14} + \nnn{16}$ & 2 & $\partial^\mu [(\covD_\mu {D^\pm}^*) (\covD_\nu D^\pm) + (\covD_\nu {D^\pm}^*) (\covD_\mu D^\pm)] \, Z^\nu$ \\
\hline
$\nnn{15} - \nnn{17}$ & 2 & $i \, \partial^\mu [(\covD_\mu {D^\pm}^*) (\covD_\nu D^\pm) - (\covD_\nu {D^\pm}^*) (\covD_\mu D^\pm)] \, Z^\nu$ \\
\hline
$\nnn{16} + \nnn{18}$ & 2 & $\partial^\mu [(\covD_\mu \covD_\nu {D^\pm}^*) D^\pm + {D^\pm}^* (\covD_\mu \covD_\nu D^\pm)] \, Z^\nu$ \\
\hline
$\nnn{17} + \nnn{19}$ & 2 & $i \, \partial^\mu [(\covD_\mu \covD_\nu {D^\pm}^*) D^\pm - {D^\pm}^* (\covD_\mu \covD_\nu D^\pm)] \, Z^\nu$ \\
\hline
$(\nnn{14} + \nnn{16}) +$ & \multirow{2}{*}{2} & \multirow{2}{*}{$\partial^2 \partial_\mu ({D^\pm}^* D^\pm) \, Z^\mu$} \\
$(\nnn{16} + \nnn{18})$ & & \\
\hline
$(\nnn{15} - \nnn{17}) -$ & \multirow{2}{*}{2} & \multirow{2}{*}{$i \, \partial^2 ({D^\pm}^* \overleftrightarrow{\covD_\mu} D^\pm) \, Z^\mu$} \\
$(\nnn{17} + \nnn{19})$ & & \\
\hline
$- (\nnn{15} - \nnn{17})$ & \multirow{3}{*}{2} & \multirow{3}{*}{$i \, \partial_\mu \partial_\nu ({D^\pm}^* \overleftrightarrow{\covD^\mu} D^\pm) \, Z^\nu$} \\
$- (\nnn{17} + \nnn{19})$ & & \\
$= - \nnn{15} - \nnn{19}$ & & \\

\hline
\end{longtable}
\end{center}

The following linear combinations are total derivatives:
\\
$\nnn{1} + \nnn{2}$ and $\nnn{1} + \nnn{3}$
\\
$\nnn{6} + \nnn{16}$
\\
$\nnn{7} + (\nnn{14} + \nnn{16})$
\\
$\nnn{12} + (\nnn{16} + \nnn{18})$
\\
$\nnn{13} + (\nnn{17} + \nnn{19})$
\\
$\nnn{2} + (\nnn{7} + \nnn{12})$ and $\nnn{3} + (\nnn{7} + \nnn{12})$ and $\nnn{3} + (2 * \nnn{6} + \nnn{10})$
\\
$(\nnn{7} + \nnn{12}) + (\nnn{14} + \nnn{16}) + (\nnn{16} + \nnn{18})$ and $(2 * \nnn{6} + \nnn{10}) + (\nnn{14} + \nnn{16}) + (\nnn{16} + \nnn{18})$
\\
$\nnn{4} + (\nnn{8} - \nnn{13})$ and $(\nnn{8} - \nnn{13}) + ((\nnn{15} - \nnn{17}) - (\nnn{17} + \nnn{19}))$
\\
$\nnn{5} - (\nnn{8} + \nnn{13})$ and $- (\nnn{8} + \nnn{13}) + (- \nnn{15} - \nnn{19})$
\\
$\nnn{5} + \nnn{11}$ and $\nnn{11} + (- \nnn{15} - \nnn{19})$.

\subsection{Terms with four $D$'s}

\subsubsection{With no derivatives}
\begin{center}
\begin{longtable}{|>{\pnt} r | c | c |}
\hline
n & $D - 4$ & term \\
\hline
& 2 & ${D^0}^* D^0 \, ({D^0}^* D^\pm W^\mp_\mu + D^0 {D^\pm}^* W^\pm_\mu) \, Z^\mu$ \\
\hline
& 2 & $i \, {D^0}^* D^0 \, ({D^0}^* D^\pm W^\mp_\mu - D^0 {D^\pm}^* W^\pm_\mu) \, Z^\mu$ \\

\doubline

& 2 & ${D^0}^* D^0 {D^\pm}^* D^\pm Z^\mu Z_\mu$ \\

\doubline

& 2 & ${D^0}^* D^0 {D^\pm}^* D^\pm W^{+\mu} W^-_\mu$ \\

\doubline

& 2 & ${D^0}^* {D^0}^* D^\pm D^\pm W^{\mp\mu} W^\mp_\mu + D^0 D^0 {D^\pm}^* {D^\pm}^* W^{\pm\mu} W^\pm_\mu$ \\
\hline
& 2 & $i \, ({D^0}^* {D^0}^* D^\pm D^\pm W^{\mp\mu} W^\mp_\mu - D^0 D^0 {D^\pm}^* {D^\pm}^* W^{\pm\mu} W^\pm_\mu)$ \\

\doubline

& 2 & ${D^\pm}^* D^\pm ({D^0}^* D^\pm W^\mp_\mu + D^0 {D^\pm}^* W^\pm_\mu) Z^\mu$ \\
\hline
& 2 & $i \, {D^\pm}^* D^\pm ({D^0}^* D^\pm W^\mp_\mu - D^0 {D^\pm}^* W^\pm_\mu) Z^\mu$ \\

\doubline

& 2 & ${D^\pm}^* D^\pm {D^\pm}^* D^\pm Z^\mu Z_\mu$ \\

\doubline

& 2 & ${D^\pm}^* D^\pm {D^\pm}^* D^\pm W^{+\mu} W^-_\mu$ \\

\hline
\end{longtable}
\end{center}

\subsubsection{With one derivative}
As in \ref{D^1 D^2} we separate the list of terms in two parts, the first with non-hermitian terms (all the hermitian terms can then be derived from these by taking their real and imaginary parts), and the second with hermitian terms.
\\
The non-hermitian terms are:
\begin{center}
\begin{longtable}{|>{\pnt} r | c | c |}
\hline
n & $D - 4$ & term \\
\hline
& 2 & $(\partial_\mu {D^0}^*) D^0 {D^0}^* D^\pm W^{\mp\mu}$ \\
\hline
& 2 & ${D^0}^* (\partial_\mu D^0) {D^0}^* D^\pm W^{\mp\mu}$ \\
\hline
& 2 & ${D^0}^* D^0 {D^0}^* (\covD_\mu D^\pm) W^{\mp\mu}$ \\

\doubline

& 2 & $(\partial_\mu {D^0}^*) D^\pm {D^\pm}^* D^\pm W^\mp_\mu$ \\
\hline
& 2 & ${D^0}^* (\covD^\mu D^\pm) {D^\pm}^* D^\pm W^\mp_\mu$ \\
\hline
& 2 & ${D^0}^* D^\pm (\covD^\mu {D^\pm}^*) D^\pm W^\mp_\mu$ \\
\hline
& 2 & ${D^0}^* D^\pm {D^\pm}^* (\covD^\mu D^\pm) W^\mp_\mu$ \\

\hline

\end{longtable}
\end{center}

The hermitian terms are:
\begin{center}
\begin{longtable}{|>{\pnt} r | c | c |}
\hline
n & $D - 4$ & term \\
\hline
& 2 & $\partial_\mu ({D^0}^* D^0) \, {D^\pm}^* D^\pm Z^\mu$ \\
\hline
& 2 & $J_\mu {D^\pm}^* D^\pm Z^\mu$ \\
\hline
& 2 & ${D^0}^* D^0 \, \partial_\mu ({D^\pm}^* D^\pm) \, Z^\mu$ \\
\hline
& 2 & $i \, {D^0}^* D^0 \, ({D^\pm}^* \overleftrightarrow{\covD_\mu} D^\pm) \, Z^\mu$ \\

\doubline

& 2 & $\partial_\mu ({D^\pm}^* D^\pm) \, {D^\pm}^* D^\pm Z^\mu$ \\
\hline
& 2 & $i \, ({D^\pm}^* \overleftrightarrow{\covD_\mu} D^\pm) \, {D^\pm}^* D^\pm Z^\mu$ \\

\hline

\end{longtable}
\end{center}
where $J_\mu = i \, {D^0}^* \overleftrightarrow{\partial_\mu} D^0$ as defined in Appendix \ref{singInt}.

\subsubsection{With two derivatives}\label{D^2 D^4}
In the following we mark with a $\spadesuit$ those operators that do not describe interactions of the doublet with the SM particles, constituting therefore doublet self-interaction terms. We list them here because they combine with other terms to yield total derivatives vanishing upon integration by parts. We refer to these combinations at the bottom of the table. All the terms appearing here are included in the list of the doublet's self-interaction terms \eqref{DoubSelfeq}.
\begin{center}
\begin{longtable}{|>{\pnt} r | c | c |}
\hline
n & $D - 4$ & term \\
\hline
$\spadesuit$ $\nnn{4}$ & 2 & $\partial^\mu ({D^0}^* D^0) \, \partial_\mu ({D^\pm}^* D^\pm)$ \\
\hline
$\nnn{5}$ & 2 & $i \, \partial^\mu ({D^0}^* D^0) \, ({D^\pm}^* \overleftrightarrow{\covD_\mu} D^\pm)$ \\
\hline
& 2 & $i \, J^\mu ({D^\pm}^* \overleftrightarrow{\covD_\mu} D^\pm)$ \\
\hline
$\nnn{7}$ & 2 & ${D^0}^* D^0 \, (\covD^\mu {D^\pm}^*) (\covD_\mu D^\pm)$ \\
\hline
$\nnn{8}$ & 2 & ${D^0}^* D^0 \, [(\covD^\mu \covD_\mu {D^\pm}^*) \, D^\pm + {D^\pm}^* \, (\covD^\mu \covD_\mu D^\pm)]$ \\
\hline
$\spadesuit$ $2 * \nnn{7} + \nnn{8}$ & 2 & ${D^0}^* D^0 \, \partial^\mu \partial_\mu ({D^\pm}^* D^\pm)$ \\
\hline
\multirow{2}{*}{$\nnn{9}$} & \multirow{2}{*}{2} & $i \, {D^0}^* D^0 \, [(\covD^\mu \covD_\mu {D^\pm}^*) \, D^\pm - {D^\pm}^* \, (\covD^\mu \covD_\mu D^\pm)]$ \\
& & $= - i \, {D^0}^* D^0 \, \partial^\mu ({D^\pm}^* \overleftrightarrow{\covD_\mu} D^\pm)$ \\

\doubline

$\nnn{21}$ & 2 & $[(\covD^\mu \covD_\mu {D^\pm}^*) D^\pm + {D^\pm}^* (\covD^\mu \covD_\mu D^\pm)] \, {D^\pm}^* D^\pm$ \\
\hline
& \multirow{2}{*}{2} & $i \, [(\covD^\mu \covD_\mu {D^\pm}^*) D^\pm - {D^\pm}^* (\covD^\mu \covD_\mu D^\pm)] \, {D^\pm}^* D^\pm$ \\
& & $= - i \, \partial^\mu ({D^\pm}^* \overleftrightarrow{\covD_\mu} D^\pm) \, {D^\pm}^* D^\pm$ \\
\hline
$\nnn{23}$ & 2 & $(\covD^\mu {D^\pm}^*) (\covD_\mu D^\pm) \, {D^\pm}^* D^\pm$ \\
\hline
$\spadesuit$ $\nnn{21} + 2 * \nnn{23}$ & 2 & $\partial^\mu \partial_\mu ({D^\pm}^* D^\pm) \, {D^\pm}^* D^\pm$ \\
\hline
& 2 & $({D^\pm}^* \overleftrightarrow{\covD^\mu} D^\pm) \, ({D^\pm}^* \overleftrightarrow{\covD_\mu} D^\pm)$ \\
\hline
\end{longtable}
\end{center}

The following linear combination is a total derivative no longer depending on the SM fields:
\\
$\nnn{4} + (2 * \nnn{7} + \nnn{8})$
\\
and therefore is a $\spadesuit$ type of operator.
We still have the following combination which is a total derivative involving the SM fields:
\\
$\nnn{5} - \nnn{9}$.

\section{Triplet's Interaction Terms with SM Gauge Bosons}\label{tripInt}

We enlarge further the list of operators given in the two previous appendices allowing the contemporary presence of a positive and a negative charged state, beside the neutral one. These three states fit in an isotriplet $\left( T^+, T^0, T^- \right)$. The full list of interaction terms between its components and the SM gauge bosons is now given by the terms presented in this section plus those listed in Appendix \ref{singInt} and \ref{doubInt}, provided one identifies the same electric charge components: $\phi = D^0 = T^0$ and $D^\pm = T^\pm$. While in the case of the doublet only one charged component was present, namely $D^+$ or $D^-$, now both electric charges appear at the same time as $T^+$ and $T^-$. Therefore in considering the terms in Appendix \ref{doubInt} both signs have to be taken into account. In the same way, whenever a $\pm$ or $\mp$ appear in the terms listed below, both the possibilities have to be considered.
\\
Like the singlet and the doublet, also the triplet is charged under a non-SM global $U(1)$ symmetry (e.g. the Technibarion number). We impose invariance under this extra abelian symmetry at the Lagrangian level. The full list of interactions of these DM particles with the SM gauge bosons is given by the operators presented in this and the two previous appendices. We consider hermitian operators preserving Lorentz invariance, electric charge and color gauge symmetry up to dimension ($D$) six in the mass. For further details and the description of the notation see Appendix \ref{singInt}.



%
%
To insure invariance under $U(1)_\text{EM}$ and $SU(3)_\text{c}$ the photon and gluon fields can only enter the Lagrangian in the form of field strength terms and covariant derivatives.

\subsection{Interaction with electroweak gauge bosons $W$, $Z$ and $A$}


\subsubsection{With no derivatives}
\begin{center}
\begin{longtable}{|>{\pnt} r | c | c |}
\hline
n & $D - 4$ & term \\
\hline
& 0 & ${T^+}^* T^- W^{+\mu} W^+_\mu + {T^-}^* T^+ W^{-\mu} W^-_\mu$ \\
\hline
& 0 & $i \, ({T^+}^* T^- W^{+\mu} W^+_\mu - {T^-}^* T^+ W^{-\mu} W^-_\mu)$ \\

\doubline

& 2 & $({T^+}^* T^- W^{+\mu} W^+_\mu + {T^-}^* T^+ W^{-\mu} W^-_\mu) \, Z^\nu Z_\nu$ \\
\hline
& 2 & $i \, ({T^+}^* T^- W^{+\mu} W^+_\mu - {T^-}^* T^+ W^{-\mu} W^-_\mu) \, Z^\nu Z_\nu$ \\
\hline
& 2 & $({T^+}^* T^- W^+_\mu W^+_\nu + {T^-}^* T^+ W^-_\mu W^-_\nu) \, Z^\mu Z^\nu$ \\
\hline
& 2 & $i \, ({T^+}^* T^- W^+_\mu W^+_\nu - {T^-}^* T^+ W^-_\mu W^-_\nu) \, Z^\mu Z^\nu$ \\

\doubline

& 2 & $({T^+}^* T^- W^{+\mu} W^+_\mu + {T^-}^* T^+ W^{-\mu} W^-_\mu) \, W^{+\nu} W^-_\nu$ \\
\hline
& 2 & $i \, ({T^+}^* T^- W^{+\mu} W^+_\mu - {T^-}^* T^+ W^{-\mu} W^-_\mu) \, W^{+\nu} W^-_\nu$ \\
\hline

\end{longtable}
\end{center}

\subsubsection{With one derivative}

\begin{center}
\begin{longtable}{|>{\pnt} r | c | c |}
\hline
n & $D - 4$ & term \\
\hline

$\nnn{14}$ & 2 & $[(\covD_\mu {T^+}^*) T^- \, W^{+\nu} W^+_\nu + {T^-}^* (\covD_\mu T^+) \, W^{-\nu} W^-_\nu] \, Z^\mu$ \\
\hline
$\nnn{14 i}$ & 2 & $i \, [(\covD_\mu {T^+}^*) T^- \, W^{+\nu} W^+_\nu - {T^-}^* (\covD_\mu T^+) \, W^{-\nu} W^-_\nu] \, Z^\mu$ \\
\hline
$\nnn{15}$ & 2 & $[(\covD_\mu {T^-}^*) T^+ \, W^{-\nu} W^-_\nu + {T^+}^* (\covD_\mu T^-) \, W^{+\nu} W^+_\nu] \, Z^\mu$ \\
\hline
$\nnn{15 i}$ & 2 & $i \, [(\covD_\mu {T^-}^*) T^+ \, W^{-\nu} W^-_\nu - {T^+}^* (\covD_\mu T^-) \, W^{+\nu} W^+_\nu] \, Z^\mu$ \\
\hline
$\nnn{14} + \nnn{15}$ & 2 & $[\covD_\mu ({T^+}^* T^-) \, W^{+\nu} W^+_\nu + \covD_\mu ({T^-}^* T^+) \, W^{-\nu} W^-_\nu] \, Z^\mu$ \\
\hline
$\nnn{15} - \nnn{14}$ & 2 & $[({T^+}^* \overleftrightarrow{\covD_\mu} T^-) \, W^{+\nu} W^+_\nu - ({T^-}^* \overleftrightarrow{\covD_\mu} T^+) \, W^{-\nu} W^-_\nu] \, Z^\mu$ \\
\hline
$- (\nnn{14 i} + \nnn{15 i})$ & 2 & $i \, [({T^+}^* \overleftrightarrow{\covD_\mu} T^-) \, W^{+\nu} W^+_\nu + ({T^-}^* \overleftrightarrow{\covD_\mu} T^+) \, W^{-\nu} W^-_\nu] \, Z^\mu$ \\
\hline
$\nnn{14 i} - \nnn{15 i}$ & 2 & $i \, [\covD_\mu ({T^+}^* T^-) \, W^{+\nu} W^+_\nu - \covD_\mu ({T^-}^* T^+) \, W^{-\nu} W^-_\nu] \, Z^\mu$ \\
\hline
& 2 & $[{T^+}^* T^- \, W^{+\nu} W^+_\nu + {T^-}^* T^+ \, W^{-\nu} W^-_\nu] \, (\partial_\mu Z^\mu)$ \\
\hline
& 2 & $i \, [{T^+}^* T^- \, W^{+\nu} W^+_\nu - {T^-}^* T^+ \, W^{-\nu} W^-_\nu] \, (\partial_\mu Z^\mu)$ \\

\doubline

$\nnn{24}$ & 2 & $[(\covD_\mu {T^+}^*) T^- \, W^{+\mu} W^+_\nu + {T^-}^* (\covD_\mu T^+) \, W^{-\mu} W^-_\nu] \, Z^\nu$ \\
\hline
$\nnn{24 i}$ & 2 & $i \, [(\covD_\mu {T^+}^*) T^- \, W^{+\mu} W^+_\nu - {T^-}^* (\covD_\mu T^+) \, W^{-\mu} W^-_\nu] \, Z^\nu$ \\
\hline
$\nnn{25}$ & 2 & $[(\covD_\mu {T^-}^*) T^+ \, W^{-\mu} W^-_\nu + {T^+}^* (\covD_\mu T^-) \, W^{+\mu} W^+_\nu] \, Z^\nu$ \\
\hline
$\nnn{25 i}$ & 2 & $i \, [(\covD_\mu {T^-}^*) T^+ \, W^{-\mu} W^-_\nu - {T^+}^* (\covD_\mu T^-) \, W^{+\mu} W^+_\nu] \, Z^\nu$ \\
\hline
$\nnn{24} + \nnn{25}$ & 2 & $[\covD_\mu ({T^+}^* T^-) \, W^{+\mu} W^+_\nu + \covD_\mu ({T^-}^* T^+) \, W^{-\mu} W^-_\nu] \, Z^\nu$ \\
\hline
$\nnn{25} - \nnn{24}$ & 2 & $[({T^+}^* \overleftrightarrow{\covD_\mu} T^-) \, W^{+\mu} W^+_\nu - ({T^-}^* \overleftrightarrow{\covD_\mu} T^+) \, W^{-\mu} W^-_\nu] \, Z^\nu$ \\
\hline
$- (\nnn{24 i} + \nnn{25 i})$ & 2 & $i \, [({T^+}^* \overleftrightarrow{\covD_\mu} T^-) \, W^{+\mu} W^+_\nu + ({T^-}^* \overleftrightarrow{\covD_\mu} T^+) \, W^{-\mu} W^-_\nu] \, Z^\nu$ \\
\hline
$\nnn{24 i} - \nnn{25 i}$ & 2 & $i \, [\covD_\mu ({T^+}^* T^-) \, W^{+\mu} W^+_\nu - \covD_\mu ({T^-}^* T^+) \, W^{-\mu} W^-_\nu] \, Z^\nu$ \\
\hline
$\nnn{31}$ & 2 & $[{T^+}^* T^- \, (\covD_\mu W^{+\mu}) W^+_\nu + {T^-}^* T^+ \, (\covD_\mu W^{-\mu}) W^-_\nu] \, Z^\nu$ \\
\hline
$\nnn{31 i}$ & 2 & $i \, [{T^+}^* T^- \, (\covD_\mu W^{+\mu}) W^+_\nu - {T^-}^* T^+ \, (\covD_\mu W^{-\mu}) W^-_\nu] \, Z^\nu$ \\
\hline
$\nnn{32}$ & 2 & $[{T^+}^* T^- \, W^{+\mu} (\covD_\mu W^+_\nu) + {T^-}^* T^+ \, W^{-\mu} (\covD_\mu W^-_\nu)] \, Z^\nu$ \\
\hline
$\nnn{32 i}$ & 2 & $i \, [{T^+}^* T^- \, W^{+\mu} (\covD_\mu W^+_\nu) - {T^-}^* T^+ \, W^{-\mu} (\covD_\mu W^-_\nu)] \, Z^\nu$ \\
\hline
$\nnn{31} + \nnn{32}$ & 2 & $[{T^+}^* T^- \, \covD_\mu (W^{+\mu} W^+_\nu) + {T^-}^* T^+ \, \covD_\mu (W^{-\mu} W^-_\nu)] \, Z^\nu$ \\
\hline
$\nnn{32} - \nnn{31}$ & 2 & $[{T^+}^* T^- \, (W^{+\mu} \overleftrightarrow{\covD_\mu} W^+_\nu) + {T^-}^* T^+ \, (W^{-\mu} \overleftrightarrow{\covD_\mu} W^-_\nu)] \, Z^\nu$ \\
\hline
$\nnn{31 i} + \nnn{32 i}$ & 2 & $i \, [{T^+}^* T^- \, \covD_\mu (W^{+\mu} W^+_\nu) - {T^-}^* T^+ \, \covD_\mu (W^{-\mu} W^-_\nu)] \, Z^\nu$ \\
\hline
$\nnn{32 i} - \nnn{31 i}$ & 2 & $i \, [{T^+}^* T^- \, (W^{+\mu} \overleftrightarrow{\covD_\mu} W^+_\nu) - {T^-}^* T^+ \, (W^{-\mu} \overleftrightarrow{\covD_\mu} W^-_\nu)] \, Z^\nu$ \\
\hline
$\nnn{33}$ & 2 & $[{T^+}^* T^- \, W^{+\mu} W^+_\nu + {T^-}^* T^+ \, W^{-\mu} W^-_\nu] \, (\partial_\mu Z^\nu)$ \\
\hline
$\nnn{33 i}$ & 2 & $i \, [{T^+}^* T^- \, W^{+\mu} W^+_\nu - {T^-}^* T^+ \, W^{-\mu} W^-_\nu] \, (\partial_\mu Z^\nu)$ \\

\doubline

& 2 & $\varepsilon^{\mu\nu\rho\sigma} [{T^+}^* T^- \, (\covD_\mu W^+_\nu) W^+_\rho + {T^-}^* T^+ \, (\covD_\mu W^-_\nu) W^-_\rho] \, Z_\sigma$ \\
\hline
& 2 & $i \, \varepsilon^{\mu\nu\rho\sigma} [{T^+}^* T^- \, (\covD_\mu W^+_\nu) W^+_\rho - {T^-}^* T^+ \, (\covD_\mu W^-_\nu) W^-_\rho] \, Z_\sigma$ \\
\hline
\end{longtable}
\end{center}
Not all these operators are independent, in fact the following linear combinations amount to total derivatives:
\\
$(\nnn{24} + \nnn{25}) + (\nnn{31} + \nnn{32}) + \nnn{33}$
\\
$(\nnn{24 i} - \nnn{25 i}) + (\nnn{31 i} + \nnn{32 i}) + \nnn{33 i}$.

\subsubsection{With two derivatives}

We use here the notation introduced in section \ref{D^2 D^2}: we list only some terms, the rest being derived from these by permitting the derivates to act on the various fields in different order. When below a term the symbol \virg{$*$} appears this means that the remaining terms obtained by simply changing position of the derivatives are not displayed here but should be taken into account. One has also to pay attention to the fact that a partial derivative turns into a covariant one when it acts on a charged field. 
\begin{center}
\begin{longtable}{|>{\pnt} r | c | c |}
\hline
n & $D - 4$ & term \\
\hline

& 2 & $(\covD^\mu \covD_\mu {T^\pm}^*) T^\mp \, W^{\pm\nu} W^\pm_\nu + \text{ h.c.}$ \\
\hline
& 2 & * \\

\doubline

& 2 & $i \, [(\covD^\mu \covD_\mu {T^\pm}^*) T^\mp \, W^{\pm\nu} W^\pm_\nu - \text{ h.c.}]$ \\
\hline
& 2 & * \\

\doubline

& 2 & $(\covD_\mu \covD_\nu {T^\pm}^*) T^\mp \, W^{\pm\mu} W^{\pm\nu} + \text{ h.c.}$ \\
\hline
& 2 & * \\

\doubline

& 2 & $i \, [(\covD_\mu \covD_\nu {T^\pm}^*) T^\mp \, W^{\pm\mu} W^{\pm\nu} - \text{ h.c.}]$ \\
\hline
& 2 & * \\

\doubline

& 2 & $\varepsilon^{\mu\nu\rho\sigma} \, (\covD_\mu {T^\pm}^*) (\covD_\nu T^\mp) \, W^\pm_\rho W^\pm_\sigma + \text{ h.c.}$ \\
\hline
& 2 & * \\

\doubline

& 2 & $i \, \varepsilon^{\mu\nu\rho\sigma} \, [(\covD_\mu {T^\pm}^*) (\covD_\nu T^\mp) \, W^\pm_\rho W^\pm_\sigma - \text{ h.c.}]$ \\
\hline
& 2 & * \\

\hline
\end{longtable}
\end{center}

\subsection{Terms with four $T$'s}

\subsubsection{With no derivatives}
\begin{center}
\begin{longtable}{|>{\pnt} r | c | c |}
\hline
n & $D - 4$ & term \\
\hline

& 2 & $({T^0}^* {T^0}^* T^\pm T^\mp + T^0 T^0 {T^\pm}^* {T^\mp}^*) \, Z^\mu Z_\mu$ \\
\hline
& 2 & $i \, ({T^0}^* {T^0}^* T^\pm T^\mp - T^0 T^0 {T^\pm}^* {T^\mp}^*) \, Z^\mu Z_\mu$ \\

\doubline

& 2 & $({T^0}^* {T^0}^* T^\pm T^\mp + T^0 T^0 {T^\pm}^* {T^\mp}^*) W^{+\mu} W^-_\mu$ \\
\hline
& 2 & $i \, ({T^0}^* {T^0}^* T^\pm T^\mp - T^0 T^0 {T^\pm}^* {T^\mp}^*) W^{+\mu} W^-_\mu$ \\

\doubline

& 2 & ${T^0}^* T^0 ({T^+}^* T^- W^{+\mu} W^+_\mu + {T^-}^* T^+ W^{-\mu} W^-_\mu)$ \\
\hline
& 2 & $i \, {T^0}^* T^0 ({T^+}^* T^- W^{+\mu} W^+_\mu - {T^-}^* T^+ W^{-\mu} W^-_\mu)$ \\

\doubline

& 2 & ${T^\pm}^* T^\pm ({T^0}^* T^\mp W^\pm_\mu + T^0 {T^\mp}^* W^\mp_\mu) Z^\mu$ \\
\hline
& 2 & $i \, {T^\pm}^* T^\pm ({T^0}^* T^\mp W^\pm_\mu - T^0 {T^\mp}^* W^\mp_\mu) Z^\mu$ \\

\doubline

& 2 & ${T^+}^* T^+ {T^-}^* T^- Z^\mu Z_\mu$ \\

\doubline

& 2 & ${T^+}^* T^+ {T^-}^* T^- W^{+\mu} W^-_\mu$ \\

\doubline

& 2 & ${T^\pm}^* T^\pm \, ({T^+}^* T^- W^{+\mu} W^+_\mu + {T^-}^* T^+ W^{-\mu} W^-_\mu)$ \\
\hline
& 2 & $i \, {T^\pm}^* T^\pm \, ({T^+}^* T^- W^{+\mu} W^+_\mu - {T^-}^* T^+ W^{-\mu} W^-_\mu)$ \\

\hline
\end{longtable}
\end{center}

\subsubsection{With one derivative}
As in \ref{D^1 D^2} we ease the notation by dividing the terms in two lists:
\begin{itemize}
\item[i)] the first one contains non-hermitian terms; each non-hermitian term gives rise to two hermitian terms, namely its real and imaginary parts. \item[ii)] in the second list all the terms are already hermitian.
\end{itemize}

 \begin{center}
\begin{longtable}{|>{\pnt} r | c | c |}
\hline
n & $D - 4$ & term \\
\hline

& 2 & $(\partial_\mu {T^0}^*) T^+ {T^0}^* T^- Z^\mu$ \\
\hline
& 2 & ${T^0}^* (\covD_\mu T^+) {T^0}^* T^- Z^\mu$ \\
\hline
& 2 & ${T^0}^* T^+ {T^0}^* (\covD_\mu T^-) Z^\mu$ \\

\doubline

& 2 & $(\partial_\mu {T^0}^*) T^\pm {T^\pm}^* T^\mp W^\pm_\mu$ \\
\hline
& 2 & ${T^0}^* (\covD^\mu T^\pm) {T^\pm}^* T^\mp W^\pm_\mu$ \\
\hline
& 2 & ${T^0}^* T^\pm (\covD^\mu {T^\pm}^*) T^\mp W^\pm_\mu$ \\
\hline
& 2 & ${T^0}^* T^\pm {T^\pm}^* (\covD^\mu T^\mp) W^\pm_\mu$ \\
\hline
\end{longtable}
\end{center}

 \begin{center}
\begin{longtable}{|>{\pnt} r | c | c |}
\hline
n & $D - 4$ & term \\
\hline

& 2 & $\partial_\mu ({T^+}^* T^+) \, {T^-}^* T^- Z^\mu$ \\
\hline
& 2 & $i \, ({T^+}^* \overleftrightarrow{\covD_\mu} T^+) \, {T^-}^* T^- Z^\mu$ \\
\hline
& 2 & ${T^+}^* T^+ \partial_\mu ({T^-}^* T^-) \, Z^\mu$ \\
\hline
& 2 & $i \, {T^+}^* T^+ ({T^-}^* \overleftrightarrow{\covD_\mu} T^-) \, Z^\mu$ \\

\hline

\end{longtable}
\end{center}

\subsubsection{With two derivatives}

Using the notation introduced in section \ref{D^2 D^4}
 we mark with a $\spadesuit$ those operators that do not actually describe interactions of the multiplet with SM particles, but only interactions between the components of the DM multiplet themselves. All the terms appearing here are included in the list of the triplet's self-interaction terms \eqref{TripSelfeq}.

\begin{center}
\begin{longtable}{|>{\pnt} r | c | c |}
\hline
n & $D - 4$ & term \\
\hline

$\nnn{14}$ & 2 & $\partial^\mu (T^0 T^0) \, (\covD_\mu {T^+}^*) {T^-}^* + \text{ h.c.}$ \\
\hline
$\nnn{14 i}$ & 2 & $i \, [\partial^\mu (T^0 T^0) \, (\covD_\mu {T^+}^*) {T^-}^* - \text{ h.c.}]$ \\
\hline
$\nnn{15}$ & 2 & $\partial^\mu (T^0 T^0) \, {T^+}^* (\covD_\mu {T^-}^*) + \text{ h.c.}$ \\
\hline
$\nnn{15 i}$ & 2 & $i \, [\partial^\mu (T^0 T^0) \, {T^+}^* (\covD_\mu {T^-}^*) - \text{ h.c.}]$ \\
\hline
$\spadesuit$ $\nnn{14} + \nnn{15}$ & 2 & $\partial^\mu (T^0 T^0) \, \partial_\mu ({T^+}^* {T^-}^*) + \text{ h.c.}$ \\
\hline
$\spadesuit$ $\nnn{14 i} + \nnn{15 i}$ & 2 & $i \, [\partial^\mu (T^0 T^0) \, \partial_\mu ({T^+}^* {T^-}^*) - \text{ h.c.}]$ \\
\hline
$\nnn{15} - \nnn{14}$ & 2 & $\partial^\mu (T^0 T^0) \, ({T^+}^* \overleftrightarrow{\covD_\mu} {T^-}^*) + \text{ h.c.}$ \\
\hline
$\nnn{15 i} - \nnn{14 i}$ & 2 & $i \, [\partial^\mu (T^0 T^0) \, ({T^+}^* \overleftrightarrow{\covD_\mu} {T^-}^*) - \text{ h.c.}]$ \\
\hline
$\nnn{17}$ & 2 & $T^0 T^0 \, (\covD^\mu {T^+}^*) (\covD_\mu {T^-}^*) + \text{ h.c.}$ \\
\hline
$\nnn{17 i}$ & 2 & $i \, [T^0 T^0 \, (\covD^\mu {T^+}^*) (\covD_\mu {T^-}^*) - \text{ h.c.}]$ \\

\doubline

$\nnn{31}$ & 2 & $[(\covD^\mu \covD_\mu {T^+}^*) T^+ + {T^+}^* (\covD^\mu \covD_\mu T^+)] \, {T^-}^* T^-$ \\
\hline
\multirow{2}{*}{$\nnn{32}$} & \multirow{2}{*}{2} & $i \, [(\covD^\mu \covD_\mu {T^+}^*) T^+ - {T^+}^* (\covD^\mu \covD_\mu T^+)] \, {T^-}^* T^-$ \\
& & $= - i \, \partial^\mu ({T^+}^* \overleftrightarrow{\covD_\mu} T^+) \, {T^-}^* T^-$ \\
\hline
$\nnn{33}$ & 2 & $(\covD^\mu {T^+}^*) (\covD_\mu T^+) \, {T^-}^* T^-$ \\
\hline
$\spadesuit$ $\nnn{31} + 2 * \nnn{33}$ & 2 & $\partial^\mu \partial_\mu ({T^+}^* T^+) \, {T^-}^* T^-$ \\
\hline
$\spadesuit$ $\nnn{34}$ & 2 & $\partial^\mu ({T^+}^* T^+) \, \partial_\mu ({T^-}^* T^-)$ \\
\hline
$\nnn{35}$ & 2 & $i \, \partial^\mu ({T^+}^* T^+) \, ({T^-}^* \overleftrightarrow{\covD_\mu} T^-)$ \\
\hline
$\nnn{36}$ & 2 & $i \, ({T^+}^* \overleftrightarrow{\covD^\mu} T^+) \, \partial_\mu ({T^-}^* T^-)$ \\
\hline
& 2 & $({T^+}^* \overleftrightarrow{\covD^\mu} T^+) \, ({T^-}^* \overleftrightarrow{\covD_\mu} T^-)$ \\
\hline
$\nnn{37}$ & 2 & ${T^+}^* T^+ \, (\covD^\mu {T^-}^*) (\covD_\mu T^-)$ \\
\hline
$\nnn{38}$ & 2 & ${T^+}^* T^+ \, [(\covD^\mu \covD_\mu {T^-}^*) \, T^- + {T^-}^* \, (\covD^\mu \covD_\mu T^-)]$ \\
\hline
$\spadesuit$ $2 * \nnn{37} + \nnn{38}$ & 2 & ${T^+}^* T^+ \, \partial^\mu \partial_\mu ({T^-}^* T^-)$ \\
\hline
\multirow{2}{*}{$\nnn{39}$} & \multirow{2}{*}{2} & $i \, {T^+}^* T^+ \, [(\covD^\mu \covD_\mu {T^-}^*) \, T^- - {T^-}^* \, (\covD^\mu \covD_\mu T^-)]$ \\
& & $= - i \, {T^+}^* T^+ \, \partial^\mu ({T^-}^* \overleftrightarrow{\covD_\mu} T^-)$ \\

\hline
\end{longtable}
\end{center}

The following combinations are total derivatives not involving SM fields and therefore $\spadesuit$ objects:
\\
$(\nnn{31} + 2 * \nnn{33}) + \nnn{34}$ 
\\
$\nnn{34} + (2 * \nnn{37} + \nnn{38})$ 
\\
These terms enjoy the relation:
\\
$(\nnn{31} + 2 * \nnn{33}) + 2 * \nnn{34} + (2 * \nnn{37} + \nnn{38}) = \partial^\mu \partial_\mu ({T^+}^* T^+ \, {T^-}^* T^-)$.
\\
\\
The operators forming total derivatives and involving SM fields are:
\\
$\nnn{35} - \nnn{39}$
\\
$\nnn{36} - \nnn{32}$.

\end{document}